\documentclass[12pt]{iopart}
\usepackage{iopams}
\usepackage{enumerate}
\usepackage{amsfonts}
\usepackage{amssymb}
\usepackage{amsthm}
\usepackage{braket}
\usepackage{mathrsfs}
\usepackage{dcolumn}
\usepackage{bm}
\usepackage{graphicx, graphics}
\usepackage{color}
\usepackage[pdftex]{hyperref}
\hypersetup{
colorlinks=true,
linkcolor=[rgb]{0.0,0.1,0.8},
citecolor=[rgb]{1.0,0.0,0.0},
filecolor=magenta,
urlcolor=[rgb]{0.0,0.3,1}
}

\theoremstyle{Theorem}
\newtheorem{prop}{Proposition}

\newcommand{\pac}[1]{ \left\{ #1 \right\} }
\newcommand{\pap}[1]{\left( #1 \right)}
\newcommand{\pas}[1]{\left[#1 \right]}

\newcommand\restr[2]{{
\left.\kern-\nulldelimiterspace 
#1 
\vphantom{\big|} 
\right|_{#2} 
}}

\begin{document}
\title[Uhlmann Fidelity \& Fidelity Susceptibility for Integrable Spin Chains]{Uhlmann Fidelity and Fidelity Susceptibility for Integrable Spin Chains at Finite Temperature: Exact Results}

\author{Micha\l{} Bia\l{}o{\'n}czyk$^{1,*}$\href{https://orcid.org/0000-0001-6909-6213}{\includegraphics[scale=0.5]{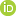}},
Fernando Javier G\'omez-Ruiz$^{2,3,4}$\href{https://orcid.org/0000-0002-1855-0671}{\includegraphics[scale=0.5]{orcid}},
and
Adolfo del Campo$^{4,2,5}$\href{https://orcid.org/0000-0003-2219-2851}{\includegraphics[scale=0.5]{orcid}}
}
\address{$^1$ Institute of Theoretical Physics, Jagiellonian University, \L{}ojasiewicza 11, 30-348 Krak\'{o}w, Poland}
\address{$^2$ Donostia International Physics Center, E-20018 San Sebasti\'an, Spain}
\address{$^3$ Departamento de F{\'i}sica, Universidad de los Andes, A.A. 4976, Bogot{\'a} D. C., Colombia}
\address{$^4$ Department of Physics and Materials Science, University of Luxembourg, L-1511 Luxembourg, Luxembourg}
\address{$^5$ Department of Physics, University of Massachusetts, Boston, MA 02125, USA}
\ead{$^{*}$\href{mailto:mbialoncz@gmail.com}{mbialoncz@gmail.com}}

\begin{abstract}
We derive the exact expression for the Uhlmann fidelity between arbitrary thermal Gibbs states of the quantum XY model in a transverse field with finite system size. Using it, we conduct a thorough analysis of the fidelity susceptibility of thermal states for the Ising model in a transverse field. We compare the exact results with a common approximation that considers only the positive-parity subspace, which is shown to be valid only at high temperatures. The proper inclusion of the odd parity subspace leads to the enhancement of maximal fidelity susceptibility in the intermediate range of temperatures. We show that this enhancement persists in the thermodynamic limit and scales quadratically with the system size. The correct low-temperature behavior is captured by an approximation involving the two lowest many-body energy eigenstates, from which simple expressions are obtained for the thermal susceptibility and specific heat.
\end{abstract}
\\
\submitto{\NJP {\bf 23}, 093033 (2021)}
\maketitle

\section{Introduction}
Quantum phase transitions, unlike classical phase transitions, are induced by quantum fluctuations related to Heisenberg uncertainty and can occur at zero temperature. They are usually triggered by a change of a macroscopic parameter, like an external magnetic field, when two competing parts of the system Hamiltonian (represented by noncommuting operator terms) exchange magnitude. The crossing of a phase transition is signaled by a profound change in ground-state properties that are reflected in the order parameter (e.g., the magnetization in a ferromagnet) and correlation functions~\cite{Amico2008,sachdev_2011,Suzuki_2013}. Quantum information theory provides a plethora of theoretical tools to analyze these properties and acquire more insights into the mechanism of quantum phase transitions~\cite{information_transitions}. Here, we focus on the fidelity that  for pure states is defined in terms of  the overlap between the ground states corresponding to close values of the external parameter $g$ that drives the phase transition:
\begin{equation}\label{fidoverlap}
F_0\pap{g_1\vert g_2} = \vert \braket{\psi(g_1) \vert \psi(g_2)} \vert,
\end{equation}
where $\ket{\psi(g)}$ denotes the ground state of parameter-dependent Hamiltonian $H(g)$.

The ground-state fidelity quantifies how sensitive the ground-state is to changes in the control parameter $g$. The potential of the ground-state fidelity as an indicator of phase transitions became apparent in the work of Quan, \emph{et al}~\cite{Quan2006}, where it was observed that the decay of Loschmidt echo is enhanced by criticality~\cite{ViyuelaPRB18}. Shortly after, Zanardi and Panukovic~\cite{Zanardi2006} proposed that static fidelity can be a good indicator of the phase transition and demonstrated this in the one-dimensional transverse field XY model and the Dicke model. These results initiated a vivid research activity focused on the use of fidelity to characterize quantum critical systems~\cite{Gu2008}. The success of this approach lies in the ability of the fidelity to capture abrupt changes in the ground state caused by a small variation of the driving parameter, with no prior knowledge of the order parameter or the location of the critical point. This observation suggests identifying the leading nontrivial term in a Taylor expansion in the shift of the control parameter
\begin{equation}
F_0\pap{g\vert g+\delta} = \vert \braket{\psi(g) \vert \psi(g+\delta)} \vert = 1 - \frac{\chi_0(g)}{2}\delta^2 +\ldots
\end{equation}
The importance of the quantity $\chi(g)$, known as the fidelity susceptibility, was first pointed out by You, \emph{et al.}~\cite{Yu2007}. More generally, one can consider the Riemann metric on the ground-state manifold spanned by varying all driving parameters, with the diagonal elements of the Riemann tensor corresponding to fidelity susceptibility, and any singularity in them indicating a phase transition~\cite{zanardi_geometric}.

The role of fidelity has been investigated in many models, including integrable spin chains like the Ising model in a transverse field and the XY models (and more general integrable spin chains), the Lipkin-Meshkov-Glick model~\cite{Scherer_2009}, and the Bose-Hubbard model~\cite{Lacki_2013}, to name some representative examples. For generic models of correlated fermions, bosons, and spin systems, the use of Monte Carlo methods has been proposed \cite{Albuquerque10,Wang15}. Likewise, the use of tensor networks has also been advanced to compute the fidelity susceptibility, in the study of many-body quantum metrology \cite{Chabuda2020}. The family of integrable spin chains is particularly useful in this context given its representation in terms of non-interacting fermions which allows for efficient computations and plenty of analytical results. This family is a good testbed for theoretical concepts and it is believed to capture properties of quantum many-body systems with long-range interactions, that are realized in laboratory~\cite{Islam2011,Zhang2017}. A phase transition in these systems is induced by varying the external transverse magnetic field (say, in the in $z$ direction) represented by the Hamiltonian term $-g \sum_{i=1}^N \sigma_i^z$, where $N$ is the system size.

Efforts to characterize the ground-state fidelity have been extended to the fidelity between thermal states~\cite{Zanardi2007,Quan2009}. The latter introduces the inverse temperature $\beta = 1/kT$ as an external control parameter, in addition to the driving parameter $g$, and requires generalizing Eq.~(\ref{fidoverlap}) to mixed states using the Uhlmann fidelity~\cite{jozsa_fidelity,QingPRE18,BrunoPRL17}
\begin{equation}
F\pap{\beta_1,g_1\vert \beta_2, g_2} =F(\rho_1,\sigma_2)= \mathrm{Tr} \pap{\sqrt{\sqrt{\rho_1} \sigma_2 \sqrt{\rho_1}}},
\end{equation}
where $\rho_1\pap{\beta_1,g_1} = \mathrm{e}^{-\beta_1 H(g_1)}/ \mathrm{Tr}(\mathrm{e}^{-\beta_1 H(g_1)})$ and $\sigma_2 \pap{\beta_2,g_2}= \mathrm{e}^{-\beta_2 H(g_2)}/ \mathrm{Tr}(\mathrm{e}^{-\beta_2 H(g_2)})$. One can define two kinds of susceptibilities,  one with respect to a change of temperature with fixed value of the driving parameter, and the second one, with respect to a change of the parameter $g$ at constant temperature:
\begin{eqnarray} \label{eq:thermal_susc}
\xi(\beta,g) &= - \frac{d^2 }{d \delta^2}F\pap{\beta,g\vert\beta+\delta,g}\Bigg|_{\delta=0},\\
\chi(\beta,g) &= - \frac{d^2 }{d \delta^2}F\pap{\beta,g\vert\beta,g+\delta}\Bigg|_{\delta=0}.\label{eq:field_susc}
\end{eqnarray}
It was shown~\cite{Quan2009} that the thermal fidelity susceptibility $\xi(\beta,g)$ is proportional to the specific heat. Further,  for a sufficiently large temperature, the fidelity susceptibility related to the field $\chi(\beta,g)$ is proportional to magnetic susceptibility, as in this limit density matrices approximately commute and the phase transition can be treated as approximately classical.

On the other hand, one can consider the limit $\beta \rightarrow \infty$, when one expects convergence to the results for ground-state fidelity. This limit justifies the expectation for the fidelity to remain a good indicator of phase transitions at finite temperature.

In this work, we focus on the exact characterization of the fidelity and fidelity susceptibility of finite-size integrable spin systems, paying particular attention to the intermediate temperature regime. Specifically, we consider that the thermal energy is comparable with the energy gap (or elementary excitation) of the system. At the critical point (and its vicinity), the fidelity susceptibility exhibits a sharp peak. The height of this peak measures how different are the states in the two phases. We provide exact formulas for the fidelity between thermal states of integrable spin chains and analyze the dependence of the height of the peak on temperature. Many spin chains widely considered in the literature share a common property, namely their Hamiltonian commutes with the parity operator
\begin{equation}
P = \prod_{i=1}^N \sigma_i^z.
\end{equation}
The latter has eigenvalues $\pm 1$ and therefore, the energy spectrum splits into two parts of positive and negative parity. For instance, in the Ising model for an even number of spins the ground state is in the positive parity subspace, but dealing with thermal states requires careful treatment of eigenstates of both parities. The necessity of proper handling of parities in finite-temperature spin chains was first stated by Katsura~\cite{Katsura62}. Later, Kapitonov and Il'inskii~\cite{Kapitonov1998} provided a derivation of full partition function using integrals over Grassmann variables. In our recent work~\cite{Bialonczyk2020}, we provide an elementary method to derive exact expressions for partition function and characteristic function of a wide class of observables using only the structure of Hilbert space. The big discrepancy, even in the thermodynamic limit, between the fidelity susceptibility of lowest energy states in positive and negative parity subspaces, was first reported by Damski and Rams~\cite{Damski2014}. A thorough analysis of the scaling of the ground state fidelity susceptibility in the XY model was conducted in~\cite{Damski_2011},  while closed-form expressions for the fidelity susceptibility in the quantum Ising model in a  transverse field were derived in \cite{Damski2013}. However, to our best knowledge, an exact treatment of the fidelity between thermal states of integrable spin chains in the complete Hilbert spaces seems to be lacking at the time of writing. We aim at filling this gap by using methods developed in~\cite{Bialonczyk2020}. We provide exact expressions for fidelity between two arbitrary thermal states in the quantum XY models and analyze the temperature dependence of fidelity susceptibility in a paradigmatic test-bed for quantum critical phenomena, the quantum  Ising model in a transverse field.

In the remaining part of the paper, we first review the diagonalization of the XY model with particular stress on the exact treatment of the parity subspaces in~\Sref{secXYdiag}. In ~\Sref{secHilbert} we recall the methods developed in~\cite{Bialonczyk2020} and use them to derive the full expression for Uhlmann fidelity between arbitrary thermal states in~\Sref{secUhlmann}. We then focus on fidelity susceptibility computed at the critical point, analyze its temperature dependence and compare it with the simplified expression obtained considering only the positive parity contribution; see~\Sref{secChi}. Although such simplification is often justified in the thermodynamic limit, we show that this is not the case for fidelity susceptibility. In particular, we show that the discrepancy between exact and simplified expressions persists when increasing the system size. We also introduce an accurate low-temperature approximation using two lowest-lying energy eigenstates, that we use to characterize the thermal susceptibility and specific heat. Our results are thus of direct relevance to the use of the fidelity and fidelity susceptibility to characterize quantum critical phenomena.

\section{Diagonalization of the XY chain}\label{secXYdiag}

In this section, we outline the basic steps to diagonalize the XY model in one spatial dimension. The method we use can be applied as well in extended XY models~\cite{ExtendedXY}, with special care of the parity of the ground state. The Hamiltonian of an XY chain is
\begin{equation}\label{eq:XYHamiltonian}
H(\gamma, g) = -\sum_{i=1}^N \pap{ \frac{1+\gamma}{2} \sigma_i^x \sigma_{i+1}^x + \frac{1-\gamma}{2} \sigma_i^y \sigma_{i+1}^y} - g\sum_{i=1}^N \sigma_i^z.
\end{equation}
When $\gamma = 1$, the Hamiltonian~\eref{eq:XYHamiltonian} corresponds to the Ising model in a transverse magnetic field (TFQIM), while the limit $\gamma = 0$ describes the isotropic XY model. For the anisotropic case $0 < \gamma \leq 1$ the model belongs to the Ising universality class, and its phase diagram is determined by the value of $g$. When $g > 1$ the magnetic field dominates over the nearest-neighbor coupling, polarizing the spins along the $z$ direction. This corresponds to a paramagnetic state, with zero magnetization in the $xy$ plane. On the other hand, when $0 \leq g < 1$ the ground state of the system corresponds to a ferromagnetic configuration with polarization along the $xy$ plane. These phases are separated by a second-order quantum phase transition (QPT) at the critical point $g = 1$. Finally, for the isotropic case $\gamma = 0$, a QPT is observed between the gapless phase ($g < 1$) and the ferromagnetic phase ($g > 1$).

In the following, we assume periodic boundary conditions, i.e.,  $\sigma_{N+1}^{\alpha}=\sigma_{1}^{\alpha}$. The Hamiltonian~\eref{eq:XYHamiltonian} can be diagonalized with mapping onto noninteracting fermions using the Jordan-Wigner transformation~\cite{Lieb1961}
\begin{equation}
\sigma_i^z = 1-2c_i^{\dagger} c_i, \quad \sigma_i^x = (c_i+c_i^{\dagger}) \prod_{j<i} (1-2 c_j^{\dagger} c_j), \quad 
\sigma_i^y = \mathrm{i} (c_i-c_i^{\dagger}) \prod_{j<i} (1-2 c_j^{\dagger} c_j),
\end{equation}
where the fermionic operators satisfy $\pac{c_i,c_j} = 0$ and $\pac{c_i,c_j^{\dagger}} = \delta_{ij}$.
Using the fermionic representation in real space, the Hamiltonian takes the form
\begin{eqnarray}\label{eq:XYFerm}
H\pap{\gamma,g}&=-\sum_{i=1}^{N-1}\pas{c_{i}^{\dagger}c_{i+1} +c_{i+1}^{\dagger}c_{i}+\gamma\pap{c_{i}^{\dagger}c_{i+1}^{\dagger} + c_{i+1}c_{i}}}\nonumber\\
&\quad+P\pas{c_{N}^{\dagger}c_{1}+c_{1}^{\dagger}c_{N}+\gamma\pap{c_{N}c_{1}+c_1c_{N}}}\\
&\quad-g\sum_{i=1}^{N}\pap{1-2c_{i}^{\dagger}c_{i}},\nonumber
\end{eqnarray}
where $P = \prod_{i=1}^N \sigma_i^z$ is the \emph{parity operator} already mentioned in the introduction. The next step is to make use of the Fourier transform
\begin{equation}
c_j = \frac{\exp\pap{-\mathrm{i} \pi/4}}{\sqrt{N}} \sum_{k \in \mathbf{K}} c_k \exp\pap{\mathrm{i} k j}.
\end{equation}
Because of the presence of the parity operator, Hamiltonian~\eref{eq:XYFerm} is not in the form of noninteracting fermions yet. However, the operator $P$ has eigenvalues $\pm 1$ and commutes with the Hamiltonian. Therefore, the Hamiltonian can be diagonalized separately in two sectors of the total Hilbert space, with $P=1$ (even number of quasiparticles) and $P=-1$ (odd number of quasiparticles). The difficulty lies in the fact that the boundary conditions obeyed by the fermionic operators depend on the sector of the Hilbert space. This changes the set of momenta $\mathbf{K}$ relevant for the Fourier transform. In the following, we write down the most important results in both subspaces. We assume an even number of particles $N$; for details see for example~\cite{Damski2014,rams2015}.\\
\\
\emph{Positive parity subspace.} The condition $P=1$ implies $c_{N+1} = -c_1$ and the set of momenta is the following:
\begin{equation}
\mathbf{K}^+ = \pac{\pm \frac{\pi}{N}, \pm \frac{3\pi}{N}, \ldots, \pm \pap{\pi - \frac{\pi}{N}}} = \mathbf{k}^+ \cup -\mathbf{k}^+,
\end{equation}
where $\mathbf{k}^+$ denotes the set of only positive momenta: $\mathbf{k}^+ =\pac{\frac{\pi}{N}, \frac{3\pi}{N}, \ldots, \pap{\pi - \frac{\pi}{N}}}$. The Hamiltonian takes the form
\begin{equation}
H(\gamma, g) = 2 \sum_{k \in \mathbf{k}^+} \pas{\pap{c_k^{\dagger} c_k - c_{-k} c_{-k}^{\dagger}}\pap{g-\cos k} + \pap{c_k^{\dagger}c_{-k}^{\dagger} + c_k c_{-k}} \gamma\sin k}.
\end{equation}
After a Bogoliubov transformation, it can be written as
\begin{equation}
H^+(\gamma, g) = \sum_{k \in \mathbf{k}^+} 2 \epsilon_k \pap{\gamma,g}\pap{\gamma_k^{\dagger} \gamma_k + \gamma_{-k}^{\dagger} \gamma_{-k}-1},
\end{equation}
where, for all $k \in \mathbf{k}^+$:
\numparts
\begin{eqnarray}
\epsilon_k \pap{\gamma,g} = \sqrt{(g-\cos k)^2 + (\gamma \sin k)^2},\label{eq:energiesp1}\\
\gamma_k = \cos \pap{\frac{\vartheta_k}{2}} c_k + \sin\pap{\frac{\vartheta_k}{2}} c_{-k}^{\dagger}, \label{eq:energiesp2}\\ 
(\sin \vartheta_k, \cos\vartheta_k) = \pap{\frac{\gamma\sin k}{\epsilon_k \pap{\gamma,g}}, \frac{g-\cos k}{\epsilon_k\pap{\gamma,g}}}.\label{eq:energiesp3}
\end{eqnarray}
\endnumparts
The lowest energy state in this subspace and its energy are respectively given by 
\numparts
\begin{eqnarray}
\ket{\gamma,g }^+ &= \prod_{k\in \mathbf{k}^+} \pas{ \cos \pap{\frac{\vartheta_k}{2}} - \sin \pap{\frac{\vartheta_k}{2}} c_k^{\dagger} c_{-k}^{\dagger}}\ket{\rm vac}.\label{state0p}\\
E_0^+\pap{\gamma,g} &= -2 \sum_{k \in \mathbf{k}^+} \epsilon_k\pap{\gamma,g},
\end{eqnarray}
where $\ket{\rm vac}$ is a state annihilated by fermionic operators $c_k$ for $k \in \mathbf{K}^+$. 
\endnumparts

\emph{Negative parity subspace.} Similarly, the condition $P=-1$ implies $c_{N+1} = c_1$ and the set of allowed momenta
\begin{equation}
\mathbf{K}^- = \pac{0, \pm \frac{2\pi}{N}, \pm\frac{4\pi}{N}, \ldots,\pm\pap{\pi-\frac{2\pi}{N} } , \pi} = \mathbf{k}^- \cup -\mathbf{k}^-\cup \pac{0,\pi}, 
\end{equation}
where the set of positive momenta different then $0, \pi$ is $\mathbf{k}^-=\pac{\frac{2\pi}{N}, \frac{4\pi}{N}, \ldots \pap{\pi-\frac{2\pi}{N} }}$. All steps of diagonalization are the same as for the positive-parity subspace, except for the fact that the modes with $0$ and $\pi$ momenta require careful treatment. We thus repeat the steps for $k \in \mathbf{k}^-$ and treat $0,\pi$ momenta separately.

The Hamiltonian takes the form
\begin{eqnarray}
H^-(\gamma, g) =& 2 \sum_{k \in \mathbf{k}^-} \left[\pap{c_k^{\dagger} c_k - c_{-k} c_{-k}^{\dagger}}\pap{g-\cos k}\right.\nonumber\\
&\quad\quad\quad+\left.\pap{c_k^{\dagger}c_{-k}^{\dagger} + c_k c_{-k}} \gamma\sin k \right]\\
&+(g-1) (c_0^{\dagger} c_0 - c_0 c_0^{\dagger}) + (g+1) (c_{\pi}^{\dagger} c_{\pi} - c_{\pi} c_{\pi}^{\dagger}).\nonumber
\end{eqnarray}
After a Bogoliubov transformation, it can be written as
\begin{eqnarray}
H^-(\gamma, g) =& \sum_{k \in \mathbf{k}^-} 2 \epsilon_k \pap{\gamma, g}\pap{\gamma_k^{\dagger} \gamma_k + \gamma_{-k}^{\dagger} \gamma_{-k}-1} \nonumber\\
&+ (g-1) (c_0^{\dagger} c_0 - c_0 c_0^{\dagger}) + (g+1) (c_{\pi}^{\dagger}c_{\pi} - c_{\pi} c_{\pi}^{\dagger}),
\end{eqnarray}
where, for all $k \in \mathbf{k}^-$ the equations~\eref{eq:energiesp1},~\eref{eq:energiesp2}, and~\eref{eq:energiesp3} take the same forms. Additionally, for $0,\pi$ momenta 
\begin{eqnarray} \label{eq:energies0pi}
\epsilon_0 = g-1, \quad \epsilon_{\pi} = g+1, \\
\vartheta_0 = 0, \quad \vartheta_{\pi} = 0.
\end{eqnarray} 
The ground state in this subspace and its energy read 
\numparts
\begin{eqnarray}
\ket{\gamma, g}^- &=c_0^{\dagger} \prod_{k\in \mathbf{k}^-} \pas{ \cos \pap{\frac{\vartheta_k}{2}} - \sin \pap{\frac{\vartheta_k}{2}} c_k^{\dagger} c_{-k}^{\dagger}}\ket{\rm vac}.\label{state0n}\\
E_0^-\pap{\gamma, g} &= -2 \sum_{k \in \mathbf{k}^-} \epsilon_k \pap{\gamma, g}-2.
\end{eqnarray}
\endnumparts
In the limit case $\gamma=1$, the ground state of the total Hamiltonian, for an even number of spins, always lies in the positive-parity subspace~\cite{Damski2014}. In the case of the general XY model, this is not always true~\cite{rams2015}. Later on, we will analyze in detail the Ising model, in which the energy gap is well defined 
\begin{equation}
\label{eq:symmetry_gap_definition}
\Delta(g) = \Delta(\gamma=1,g) = E_0^-(g) - E_0^+(g) \geq 0. 
\end{equation}
In the following, the explicit expression for this gap at the critical point ($g=1$) will be useful \cite{Damski2014}
\begin{equation}
\label{eq:symmetry_breaking_gap}
\Delta(g=1) = 2 \tan\pap{\frac{\pi}{4N}} \approx \frac{\pi}{2N}.
\end{equation}
In the ferromagnetic phase the gap $\Delta(g)$ vanishes exponentially  with the system size and we refer to it as a ``symmetry-breaking gap'' to distinguish it from the ``dynamical gap'', which is defined as the lowest excitation within the positive-parity subspace \cite{BialonczykSymmetry}:
\begin{equation}
\mathrm{gap}(g) = \mathrm{gap}(\gamma = 1, g) = \sqrt{g^2-2g\cos(\pi/N)+1}.
\end{equation}
At the critical point, this gap behaves as
\begin{equation}
\mathrm{gap}(g=1) \approx \frac{4 \pi}{N}. 
\end{equation}
Therefore, at the critical point, the dynamical gap is approximately eight times bigger than the symmetry-breaking gap. By contrast, in the ferromagnetic phase ($g<1$) the symmetry-breaking gap is negligible in comparison with the dynamical gap. As we shall see, this fact is important in approximating the Gibbs state. The importance of the symmetry-breaking gap and its dependence on the boundary conditions was first highlighted in \cite{Carbera1987}. 

Calculations in the positive-parity subspace are sufficient at zero temperature. Moreover, the positive parity is preserved during dynamics. However, at finite-temperature both positive and negative parity subspaces play a role, given their contribution to the unnormalized Gibbs state
\begin{equation}
\tilde\rho_{\rm Gibbs} (\beta, \gamma, g) = \exp\pas{-\beta H(\gamma, g)}.
\end{equation} 
Here and elsewhere, we use a tilde to denote unnormalized density matrices.

\section{Structure of the Hilbert space}\label{secHilbert}
In this section, we briefly recall the methods and results from~\cite{Bialonczyk2020}.
To begin, we note that from the transformation to fermionic quasiparticles, the total Hilbert space $\mathcal{H}$ can be written as a tensor product of 4-dimensional Hilbert subspaces corresponding to each pair of momenta
\begin{equation}
\mathcal{H} = \bigotimes_{k \in \mathbf{k}^+} \mathcal{H}_k, \quad \mathcal{H}_k = \mathrm{span}\pac{\ket{0}_k, c_k^{\dagger} \ket{0}_k, c_{-k}^{\dagger} \ket{0}_k, c_k^{\dagger} c_{-k}^{\dagger} \ket{0}_k},
\end{equation}
where $\ket{0}_k$ is annihilated by $c_k$ and $c_{-k}$. Note that dimensions match: there are $N/2$ momenta, which gives  the total $2^N$ dimension. For negative subspace one has $0$ and $\pi$ momenta with corresponding 2-dimensional Hilbert spaces span by $\pac{\ket{0}_0, c_0^{\dagger} \ket{0}_0}$ and $\pac{\ket{0}_{\pi}, c_{\pi}^{\dagger} \ket{0}_{\pi}}$, together with  $(N/2)-1$ momenta in $\mathbf{k}^-$. On the other hand, the subspaces of the given parity have dimension two times smaller. Therefore, it is clear that usual tensor product is not adapted for manipulations involving  a definite parity. In order to handle this, we introduced the operations of ``positive'' and ``negative'' tensor products which pick only vectors of correct parity. They can be defined recursively:
\begin{eqnarray}
\mathcal{P}\pap{\mathcal{H}_{k_1}} = \mathcal{H}_{k_1}^{(p)}, \quad \mathcal{N}\pap{\mathcal{H}_{k_1}} = \mathcal{H}_{k_1}^{(n)},
\end{eqnarray}
\begin{eqnarray}
\mathcal{P}\pap{\bigotimes_{i=1}^{n+1} \mathcal{H}_{k_i}} = \mathcal{P}\pap{\bigotimes_{i=1}^{n} \mathcal{H}_{k_i}} \otimes \mathcal{H}_{k_{n+1}}^{(p)} \oplus \mathcal{N}\pap{\bigotimes_{i=1}^{n} \mathcal{H}_{k_i}}\otimes \mathcal{H}_{k_{n+1}}^{(n)} , \: n \geq 1 \nonumber\\
\mathcal{N}\pap{\bigotimes_{i=1}^{n+1} \mathcal{H}_{k_i}} = \mathcal{P}\pap{\bigotimes_{i=1}^{n} \mathcal{H}_{k_i}} \otimes \mathcal{H}_{k_{n+1}}^{(n)} \oplus \mathcal{N}\pap{\bigotimes_{i=1}^{n} \mathcal{H}_{k_i}}\otimes \mathcal{H}_{k_{n+1}}^{(p)}, \: n \geq 1\nonumber
\end{eqnarray}
where $\mathcal{H}_k^{(p)}$ and $\mathcal{H}_k^{(n)}$ are subspaces of $\mathcal{H}_k$ spanned by all positive parity an negative parity vectors:
\begin{eqnarray*}
\mathcal{H}_k^{(p)} = \mathrm{span} \pac{\ket{0}_k, c_k^{\dagger} c_{-k}^{\dagger} \ket{0}_k}, \quad \mathcal{H}_k^{(n)}=\mathrm{span} \pac{c_k^{\dagger} \ket{0}_k, c_{-k}^{\dagger} \ket{0}_k}, \\
\mathcal{H}_0^{(p)} = \mathrm{span} \pac{\ket{0}_0}, \quad \mathcal{H}_0^{(n)} = \mathrm{span} \pac{c_0^{\dagger} \ket{0}_0},\\
\mathcal{H}_{\pi}^{(p)} = \mathrm{span} \pac{\ket{0}_{\pi}}, \quad 
\mathcal{H}_{\pi}^{(n)} = \mathrm{span} \pac{c_{\pi}^{\dagger}\ket{0}_{\pi}}.
\end{eqnarray*} 
Similarly, one can define analogous relations for positive and negative parity part of a tensor product of operators. Let us assume that each $O_k$ has zero matrix elements between vectors of different parity (such as   $\bra{0_k} c_k O_k \ket{0}_k$ or $ \bra{0_k} c_k O_k c_k^{\dagger} c_{-k}^{\dagger} \ket{0}_k$ for instance). For such operators we define a positive and negative part in an intuitive way through the following relations:
\begin{eqnarray*}
\restr{O_k^{(p)}}{\mathcal{H}_k^{(p)} }= \restr{O_k}{\mathcal{H}_k^{(p)}}, \quad 
\restr{O_k^{(p)}}{\mathcal{H}_k^{(n)} }= 0, \\
\restr{O_k^{(n)}}{\mathcal{H}_k^{(n)} }= \restr{O_k}{\mathcal{H}_k^{(n)}}, \quad \restr{O_k^{(n)}}{\mathcal{H}_k^{(p)} }= 0,
\end{eqnarray*}
with recursive relations taking the form
\begin{equation}\label{eq:oprecursion1}
\mathcal{P} \left (O_{k_1} \right) = O_{k_1}^{(p)}, \quad \mathcal{N} \left (O_{k_1} \right) = \mathcal{O}_{k_1}^{(n)},
\end{equation}
\begin{equation*} 
\mathcal{P} \left ( \bigotimes_{i=1}^{n+1} O_{k_i} \right ) = \mathcal{P} \left ( \bigotimes_{i=1}^{n} O_{k_i} \right ) \otimes O_{k_{n+1}}^{(p)} + \mathcal{N} \left ( \bigotimes_{i=1}^{n} O_{k_i} \right ) \otimes O_{k_{n+1}}^{(n)}, \, \, \, n \geq 1,
\end{equation*}
\begin{equation*}
\mathcal{N} \left ( \bigotimes_{i=1}^{n+1} \mathcal{O}_{k_i} \right ) = \mathcal{N} \left ( \bigotimes_{i=1}^{n} O_{k_i} \right ) \otimes O_{k_{n+1}}^{(p)} + \mathcal{P} \left ( \bigotimes_{i=1}^{n} O_{k_i} \right ) \otimes O_{k_{n+1}}^{(n)}, \, \, \, n \geq 1. 
\end{equation*}
These relations are true provided that momenta $k_1, \ldots k_n$ are all relevant for one subspace (positive or negative parity). At this point we emphasize that the condition of vanishing ``mixing'' matrix elements is an important restriction that does not apply to many important properties, such as the longitudinal magnetizations $\sigma_i^x$ or $\sigma_i^y$. Dealing with such operators is particularly difficult because subspaces with different momenta mix. For some methods and results involving ground states, see for example~\cite{ningWu2020}. We shall make use of the following propositions:
\begin{prop}\label{prop:products}
For every operators $O_k$ and $R_k$:
\begin{equation}
\mathcal{P}\left(\bigotimes_{i=1}^n O_{k_i} \right)\, \mathcal{P}\left(\bigotimes_{i=1}^n R_{k_i} \right) = \mathcal{P}\left(\bigotimes_{i=1}^n O_{k_i} \, R_{k_i} \right) ,
\end{equation}
\begin{equation}
\mathcal{N}\left(\bigotimes_{i=1}^n O_{k_i} \right)\, \mathcal{N}\left(\bigotimes_{i=1}^n R_{k_i} \right) = \mathcal{N}\left(\bigotimes_{i=1}^n O_{k_i} \, R_{k_i} \right).
\end{equation}
\end{prop}
\begin{prop}
\label{prop:exponents}
The exponentials  of restricted tensor products obey:
\begin{equation}
\exp\mathcal{P}\left(\bigotimes_{i=1}^n O_{k_i} \right) = \mathcal{P}\left( \bigotimes_{i=1}^n \exp O_{k_i}\right),
\end{equation}
\begin{equation}
\exp\mathcal{N}\left(\bigotimes_{i=1}^n O_{k_i} \right) = \mathcal{N}\left( \bigotimes_{i=1}^n \exp O_{k_i}\right).
\end{equation}
\end{prop}
\begin{prop}
\label{prop:traces}
Traces of restricted tensor products read:
\begin{equation}
\label{eq:trace1}
\mathrm{Tr} \, \mathcal{P}\left(\bigotimes_{i=1}^n O_{k_i} \right) = \frac{1}{2} \left (\prod_{i=1}^n \mathrm{Tr} \, O_{k_i} + \prod_{i=1}^n (\mathrm{Tr} \, O_{k_i}^{(p)} - \mathrm{Tr} \, O_{k_i}^{(n)} )\right),
\end{equation}
\begin{equation}
\label{eq:trace2}
\mathrm{Tr} \, \mathcal{N}\left(\bigotimes_{i=1}^n O_{k_i} \right) = \frac{1}{2} \left (\prod_{i=1}^n \mathrm{Tr} \, O_{k_i} - \prod_{i=1}^n (\mathrm{Tr} \, O_{k_i}^{(p)} - \mathrm{Tr} \, O_{k_i}^{(n)}) \right).
\end{equation}
\end{prop} 
These propositions suffice to derive the formula for fidelity between arbitrary thermal states. We finish this section with an example:
\section*{Example 1: Canonical  Gibbs state}
The complete Hamiltonian can be written in the form:
\begin{equation}
H(\gamma,g) = H^+(\gamma,g) \oplus H^-(\gamma, g),
\end{equation}
where 
\begin{equation}
H^+(\gamma,g) = \mathcal{P}\pap{\bigotimes_{k\in \mathbf{k}^+} H_k(\gamma,g)}, \: H^-(\gamma,g) = \mathcal{N}\pap{\bigotimes_{k\in \mathbf{k}^-\cup \pac{0,\pi}} H_k(\gamma,g)}.
\end{equation}
Operators $H_k$ have the following matrix representations 
\begin{eqnarray}
H_k (\gamma,g)&= 2\left(
\begin{array}{cccc}
\cos k - g & \gamma\sin k & 0 & 0 \\
\gamma\sin k & g-\cos k & 0 & 0 \\
0 & 0 & 0& 0 \\
0 & 0 & 0 & 0 \\
\end{array}
\right),\\
H_0 (g)&= \left(
\begin{array}{cc}
1 - g & 0 \\
0 & g-1 \\
\end{array}
\right),\\
H_{\pi} (g)&= \left(
\begin{array}{cc}
-1 - g & 0 \\
0 & 1+g \\
\end{array}
\right),
\end{eqnarray}
in the basis $\pac{\ket{0}_k, c_k^{\dagger} c_{-k}^{\dagger} \ket{0}_k, c_k^{\dagger} \ket{0}_k, c_{-k}^{\dagger} \ket{0}_k}$ and $\pac{\ket{0}_{0,\pi}, c_{0,\pi}^{\dagger} \ket{0}_{0,\pi}}$, respectively.
Using Proposition~\ref{prop:exponents}, we can write the unnormalized thermal Gibbs state as:
\begin{equation}
\label{eq:gibbs_decomposition}
\tilde{\rho}_{\rm Gibbs} (\beta, \gamma, g) =\exp\left(-\beta H(\gamma, g)\right)= \mathcal{P}\pap{\bigotimes_{k\in \mathbf{k}^+} \tilde{\rho_k}} \oplus
\mathcal{N}\pap{\bigotimes_{k\in \mathbf{k}^-\cup \pac{0,\pi}} \tilde{\rho_k}},
\end{equation}
where
\begin{equation}
\tilde{\rho_k} = 
\mathrm{exp}\pas{-2\beta\left(\begin{array}{cc}
\cos k -g & \gamma\sin k \\
\gamma\sin k & g-\cos k \\
\end{array}
\right)} \oplus \mathbb{I}_2,
\quad
\end{equation}
\begin{equation}
\tilde\rho_0 = 
\left(\begin{array}{cc}
\mathrm{exp}\pas{\beta(g-1)} & 0 \\
0 & \exp{\pas{\beta(1-g)}} \\
\end{array}
\right),
\end{equation}
\begin{equation}
\tilde\rho_{\pi} = 
\left(\begin{array}{cc}
\mathrm{exp}\pas{-\beta(g+1)} & 0 \\
0 & \exp{\pas{\beta(g+1)}} \\
\end{array}
\right).
\end{equation}
The even and odd parity parts of $\rho_k$ are 
\begin{equation}
\tilde\rho_k^{(p)} = 
\mathrm{exp}\pas{-2\beta\left(\begin{array}{cc}
\cos k -g & \gamma\sin k \\
\gamma\sin k & g-\cos k \\
\end{array}
\right)}, 
\end{equation}
\begin{equation} 
\tilde\rho_k^{(n)} = \mathbb{I}_2, \quad \tilde\rho_0^{(p)} = \exp\pas{\beta(g-1)}, \quad \tilde\rho_{0}^{(n)} = \exp\pas{\beta(1-g)}.
\end{equation}
Using~\Eref{eq:energiesp1} one has $\mathrm{Tr}\pap{ \tilde\rho_k }= 2 \cosh\pap{2\beta \epsilon_k}$. Thanks to Proposition~\ref{prop:traces}, we can easily write down the formula for the partition function, that is the trace of the unnormalized Gibbs state:
\begin{eqnarray}
\label{eq:FullZ}
Z\pap{\beta,\gamma,g} &= \mathrm{Tr}\pap{ \tilde{\rho}_{\rm Gibbs}(\beta, \gamma, g)}\nonumber\\
&=\frac{1}{2}\Bigg[ \prod_{k\in \mathbf{K}^{+}} 2 \cosh\pap{\beta \epsilon_k\pap{\gamma,g}}+\prod_{k\in \mathbf{K}^{+}} 2 \sinh\pap{\beta \epsilon_k\pap{\gamma,g}}\\
&\quad+\prod_{k\in \mathbf{K}^{-}} 2 \cosh\pap{\beta \epsilon_k\pap{\gamma,g}}-\prod_{k\in \mathbf{K}^{-}} 2 \sinh\pap{\beta \epsilon_k\pap{\gamma,g}}\Bigg].\nonumber
\end{eqnarray}
Here, the use of the full sets of momenta $\mathbf{K}^+$ and $\mathbf{K}^-$ leads to a compact expression, but we emphasize the need to use the correct formulas for excitations in the modes with $0$ and $\pi$ momenta~\eref{eq:energies0pi}.

\Eref{eq:FullZ} can be divided into four parts~\cite{Kapitonov1998} $[$Positive Fermionic $\pap{Z_F^+(\beta, \gamma, g)}$, Positive Boundary $\pap{Z_B^+(\beta, \gamma, g)}$, Negative Fermionic $\pap{Z_F^-(\beta, \gamma, g)}$, and Negative Boundary $\pap{Z_F^+(\beta, \gamma, g)}]$:
\begin{equation}
\label{eq:Z_decomp}
Z(\beta, \gamma, g) = \frac{1}{2} \pas{Z_F^+(\beta, \gamma, g) + Z_B^+(\beta,\gamma, g) + Z_F^-(\beta,\gamma, g)-Z_B^-(\beta,\gamma, g)}.
\end{equation}
These four terms originate from the trace formulas in Proposition~\ref{prop:traces}. It is thus reasonable to assume that many physically relevant quantities (such as characteristic functions of observables or fidelity susceptibility) can be decomposed in an analogous form. It has been argued that in the thermodynamic limit only $Z_F^+$ is relevant, i.e.,  that the partition function is governed by the Positive Fermionic part corresponding to the first part of~\Eref{eq:trace1} ~\cite{Katsura62}. While this is true in many cases, one has to be careful; see~\cite{Bialonczyk2020} for details. Explicitly, the Positive Fermionic part of the partition function has the  form
\begin{equation}\label{eq:FermionicZ}
Z_F(\beta, \gamma, g) = \prod_{k \in \mathbf{K}^+} 2 \cosh\pap{\beta \epsilon_k\pap{\gamma,g}}. 
\end{equation}
We note that decompositions resembling~\eref{eq:Z_decomp} appear naturally in equations describing fidelities. In what follows, we use the acronym PPA to refer to the Positive Parity Approximation that includes only Positive Fermionic contribution. As we shall see, in the study of the fidelity susceptibility the exact expression involving all contributions is significantly different from the PPA approximation frequently used in the literature~\cite{sachdev_2011,Suzuki_2013,Zanardi2006,Zanardi2007,Quan2009,Pfeuty70, Barouch1975, Deng2011}.

The algebraic method we have presented for the exact treatment of the partition function is natural and elementary, but not the only one. Alternative derivations are possible making use of Grassmann variables, see~\cite{Kapitonov1998} or group theory~\cite{Fei19}.

\section{Expressions for fidelity between arbitrary thermal states}\label{secUhlmann}
In this section, we derive the exact expression for the Uhlmann fidelity between thermal states in the XY model with periodic boundary conditions. The method can be generalized to states that are exponentials or quadratic expressions in Fermionic operators.\\ 
\\
We begin by restating the definition of the Uhlmann fidelity between two density matrices $\rho$ and $\sigma$ ~\cite{Uhlmann76}
\begin{equation}\label{UhlmannEq1}
F\pap{\rho, \sigma} = \mathrm{Tr} \pap{\sqrt{\sqrt{\rho} \sigma \sqrt{\rho}}}. 
\end{equation}
As  $\sqrt{\rho}_k$ is the square root of $\rho_k$, it follows that 
\begin{equation}
\pas{\mathcal{P} \pap{\bigotimes_k \sqrt{\tilde\rho_k}}}^2 = \mathcal{P}\pap{\bigotimes_k\sqrt{\tilde\rho_k}\sqrt{\tilde\rho_k}} = \mathcal{P}\pap{\bigotimes_k\tilde\rho_k}.
\end{equation}
As a result, one finds
\begin{equation}\label{eq:positive_square}
\sqrt{\mathcal{P}\pap{\bigotimes_k\tilde\rho_k}} = \mathcal{P} \pap{\bigotimes_k \sqrt{\tilde\rho_k}},
\end{equation}
and similarly for $\sqrt{\mathcal{N}\pap{\bigotimes_k\tilde\rho_k}}$. Using the results from Example 1  and the unnormalized thermal states $\tilde{\sigma}_{\rm Gibbs}$, $\tilde{\rho}_{\rm Gibbs}$,
\begin{equation}
\tilde\rho_{\rm Gibbs} = \tilde\rho_{\rm Gibbs}^+ \oplus \tilde \rho_{\rm Gibbs}^-, \quad \sqrt{\tilde\rho_{\rm Gibbs}} = \sqrt{\tilde\rho_{\rm Gibbs}^+} \oplus \sqrt{\tilde\rho_{\rm Gibbs}^-},
\end{equation}
which yields a decomposition of the exact fidelity into the sum of the components with positive and negative parity 
\begin{eqnarray}
F\pap{\tilde{\rho}_{\rm Gibbs} , \tilde{\sigma}_{\rm Gibbs}} = F^+\pap{\tilde{\rho}_{\rm Gibbs} ,\tilde{\sigma}_{\rm Gibbs}} + F^-\pap{\tilde{\rho}_{\rm Gibbs} , \tilde{\sigma}_{\rm Gibbs} },
\end{eqnarray}
Using~\Eref{eq:positive_square} and Proposition~\ref{prop:products} again, we find
\begin{eqnarray}
F^+\pap{\tilde{\rho}_{\rm Gibbs}, \tilde{\sigma}_{\rm Gibbs}} = \mathrm{Tr}\pap{\mathcal{P}\pap{\bigotimes_{k \in \mathbf{k}^+} \sqrt{\sqrt{\tilde\rho_k} \tilde\sigma_k \sqrt{\tilde\rho_k}}}}\\
F^-\pap{\tilde{\rho}_{\rm Gibbs}, \tilde{\sigma}_{\rm Gibbs}} = \mathrm{Tr}\pap{\mathcal{N}\pap{\bigotimes_{k \in \mathbf{k}^- \cup \pac{0,\pi}} \sqrt{\sqrt{\tilde\rho_k} \tilde\sigma_k \sqrt{\tilde\rho_k}}}}.
\end{eqnarray}
Next we  make use of Proposition~\ref{prop:traces} to write down explicit expressions for general thermal states, with different temperatures and field values. We consider the fidelity between two thermal states characterized by field values $g^{\rho}, g^{\sigma}$, anisotropy coefficients $\gamma^{\rho}, \gamma^{\sigma}$ and inverse temperature $\beta^{\rho}, \beta^{\sigma}$, that is: 
\begin{eqnarray}
\tilde\rho_k = \exp \left[-2\beta^{\rho} 
\left(
\begin{array}{cc}
\cos k - g^{\rho} & \gamma^{\rho} \sin k \\
\gamma^{\rho} \sin k & g^{\rho} - \cos k \\
\end{array}
\right)\right]
\oplus \mathbb{I}_2 = \tilde\rho_k^{(p)} \oplus \mathbb{I}_2, \\
\tilde\sigma_k = 
\exp \left[-2\beta^{\sigma} 
\left(
\begin{array}{cc}
\cos k - g^{\sigma} & \gamma^{\sigma} \sin k \\
\gamma^{\sigma} \sin k & g^{\sigma}- \cos k \\
\end{array}
\right)\right]
\oplus \mathbb{I}_2 = \tilde\sigma_k^{(p)} \oplus \mathbb{I}_2.
\end{eqnarray}
Calculation of $\mathrm{Tr} \pap{\sqrt{\sqrt{\rho_k} \sigma_k \sqrt{\rho_k}}}$ for thermal states boils down to calculation of fidelity between qubit states, which has a particularly simple form~\cite{jozsa_fidelity}:
\begin{equation}
\mathrm{Tr} \pap{\sqrt{\sqrt{\tilde\rho_k^{(p)}}\tilde\sigma_k^{(p)} \sqrt{\tilde\rho_k^{(p)}}}} = \pas{\mathrm{Tr} \pap{\tilde\rho_k^{(p)} \tilde\sigma_k^{(p)) }}+ 2\,\mathrm{det} \pap{\tilde\rho_k^{(p)} \tilde\sigma_k^{(p)}}}^{1/2}.
\end{equation}
Using the Bogoliubov energy and angles for the states $\rho_k$ and $\sigma_k$, we can derive compact expressions for fidelity of thermal states. Although the overall structure of the formulas is simple, there are many parameters and expressions become lengthy. For clarity we introduce the following notation:
\begin{eqnarray}
u_k\pap{\rho\vert \sigma} = 2 \cosh\pap{\frac{\beta^{\rho}\epsilon_k^{\rho}+\beta^{\sigma}\epsilon_k^{\sigma}}{2}} \cos\pap{\frac{\vartheta_k^{\rho}-\vartheta_k^{\sigma}}{2}}, \\
v_k\pap{\rho\vert \sigma} = 2 \cosh \pap{\frac{\beta^{\rho}\epsilon_k^{\rho}-\beta^{\sigma}\epsilon_k^{\sigma}}{2}} \sin\pap{\frac{\vartheta_k^{\rho}-\vartheta_k^{\sigma}}{2}}.
\end{eqnarray}
The Uhlmann fidelity between thermal states $\rho_{\rm Gibbs}$ and $\sigma_{\rm Gibbs}$, respectively characterized  by the parameters $\beta^{\rho}, g^{\rho}, \gamma^{\rho}$ and $\beta^{\sigma}, g^{\sigma}, \gamma^{\sigma}$, reads
\begin{equation}\label{eq:fidelity_full1}
F\pap{\rho_{\rm Gibbs}, \sigma_{\rm Gibbs}} = \frac{ F^+\pap{\tilde{\rho}_{\rm Gibbs}, \tilde{\sigma}_{\rm Gibbs}} + F^-\pap{\tilde{\rho}_{\rm Gibbs}, \tilde{\sigma}_{\rm Gibbs}}}{\sqrt{Z(\beta^{\rho}, g^{\rho}, \gamma^{\rho})Z(\beta^{\sigma}, g^{\sigma}, \gamma^{\sigma})}}.
\end{equation} 
Here, the positive part  equals
\begin{eqnarray} \label{eq:fidelity_full2}
F^+\pap{\tilde{\rho}_{\rm Gibbs}, \tilde{\sigma}_{\rm Gibbs}} = \frac{1}{2}\Bigg[ &\prod_{k \in \mathbf{k}^+}\pap{ \sqrt{u_k^2\pap{\rho\vert \sigma}+v_k^2\pap{\rho\vert \sigma}}+ 2}\nonumber\\
&+\prod_{k \in \mathbf{k}^+}\pap{ \sqrt{u_k^2\pap{\rho\vert \sigma}+v_k^2\pap{\rho\vert \sigma}}-2}\Bigg],
\end{eqnarray} 
and the negative part is given by
\begin{eqnarray}\label{eq:fidelity_full3}
F^-\pap{\tilde{\rho}_{\rm Gibbs}, \tilde{\sigma}_{\rm Gibbs}}=\frac{1}{2}\Bigg[ &4\cosh\pap{\frac{\beta^{\rho} \epsilon_0^{\rho} + \beta^{\sigma} \epsilon_0^{\sigma}}{2}} \cosh\pap{\frac{\beta^{\rho} \epsilon_{\pi}^{\rho} + \beta^{\sigma} \epsilon_{\pi}^{\sigma}}{2}}\nonumber\\
&\quad\prod_{k \in \mathbf{k}^-}\pap{ \sqrt{u_k^2\pap{\rho\vert \sigma}+v_k^2\pap{\rho\vert \sigma}}+2}\\
&-
4 \sinh\pap{\frac{\beta^{\rho} \epsilon_0^{\rho} + \beta^{\sigma} \epsilon_0^{\sigma}}{2}} \sinh\pap{\frac{\beta^{\rho} \epsilon_{\pi}^{\rho} + \beta^{\sigma} \epsilon_{\pi}^{\sigma}}{2}}\nonumber\\
&\quad\prod_{k \in \mathbf{k}^-}\pap{ \sqrt{u_k^2\pap{\rho\vert \sigma}+v_k^2\pap{\rho\vert \sigma}}-2}\Bigg],\nonumber
\end{eqnarray} 
where excitations $\epsilon_k, \epsilon_{0,\pi}$ and Bogoliubov angles $\vartheta_k$ can be calculated with standard formulas~\eref{eq:energiesp1},~\eref{eq:energiesp2},~\eref{eq:energiesp3} and the exact partition function $Z$ is given by~\Eref{eq:FullZ}.

As in the case of partition function, we can single out the Positive Fermionic part
\begin{equation} \label{eq:fidelity_simp}
F^+_F\pap{\rho_{\rm Gibbs}, \sigma_{\rm Gibbs}} = \frac{\prod_{k \in \mathbf{k}^+}\pap{ \sqrt{u_k^2\pap{\rho\vert \sigma}+v_k^2\pap{\rho\vert \sigma}}+ 2}}{\sqrt{Z_F^+(\beta^{\rho}, \gamma^{\rho},g^{\rho})Z_F^+(\beta^{\sigma}, \gamma^{\sigma},g^{\sigma})}} \quad \mathrm{(PPA)} ,
\end{equation} 
where $Z_F^+$ is given by the first term in decomposition~\eref{eq:Z_decomp} and explicitly reads
\begin{equation}
Z_F^+ \pap{\beta,\gamma,g}= \prod_{k \in \mathbf{K}^+} 2 \cosh(\beta \epsilon_k\pap{\gamma,g}) \quad \mathrm{(PPA)}.
\end{equation}

The PPA formula~\eref{eq:fidelity_simp} has been commonly used in literature~\cite{Zanardi2007,Quan2009}. In the following sections, we are going to show the limits of this approximation by comparing it to the exact formula for the fidelity. As it turns out, the latter includes important physics in the intermediate temperature regime, which is missed by the simplified expression~\eref{eq:fidelity_simp}. We will use the notation 
\begin{equation}
F\pap{\rho_{\rm Gibbs}, \sigma_{\rm Gibbs}} = F\pap{\beta^{\rho}, \gamma^{\rho}, g^{\rho}\vert \beta^{\sigma}, \gamma^{\sigma}, g^{\sigma}}.
\end{equation}
There are two limiting cases of these general expressions that are particularly relevant. One concerns the Uhlmann fidelity between two thermal states at equal inverse temperature but a different value of the external control parameter $g$. This limit is natural to quantify the role of thermal excitations above the ground state. The second case concerns the Uhlmann fidelity between two thermal states with common control parameter $g$ but different inverse temperatures.  The explicit form of the Uhlmann fidelity can be easily derived in this case given that the two density matrices commute, i.e., $[\rho_{\rm Gibbs}, \sigma_{\rm Gibbs}]=0$.
As a result, the Uhlmann fidelity can be written down in terms of the partition function
\begin{equation}\label{UFcomm}
F\pap{\beta, \gamma, g\vert \beta', \gamma, g}=\frac{Z\left(\frac{\beta+\beta'}{2}, \gamma, g\right)}{\sqrt{Z(\beta, \gamma, g)Z(\beta', \gamma, g) }}.
\end{equation}
By using the full expression for the partition function~\eref{eq:Z_decomp}, one obtains the exact result, without the need to resort to the PPA based on the simplified partition function in the even parity subspace, previously considered, e.g., in~\cite{sachdev_2011,Suzuki_2013,Quan2006,Yu2007,Zanardi2007,Quan2009,Pfeuty70}.

\begin{figure}[b!]
        \centering
	\includegraphics[scale=0.75]{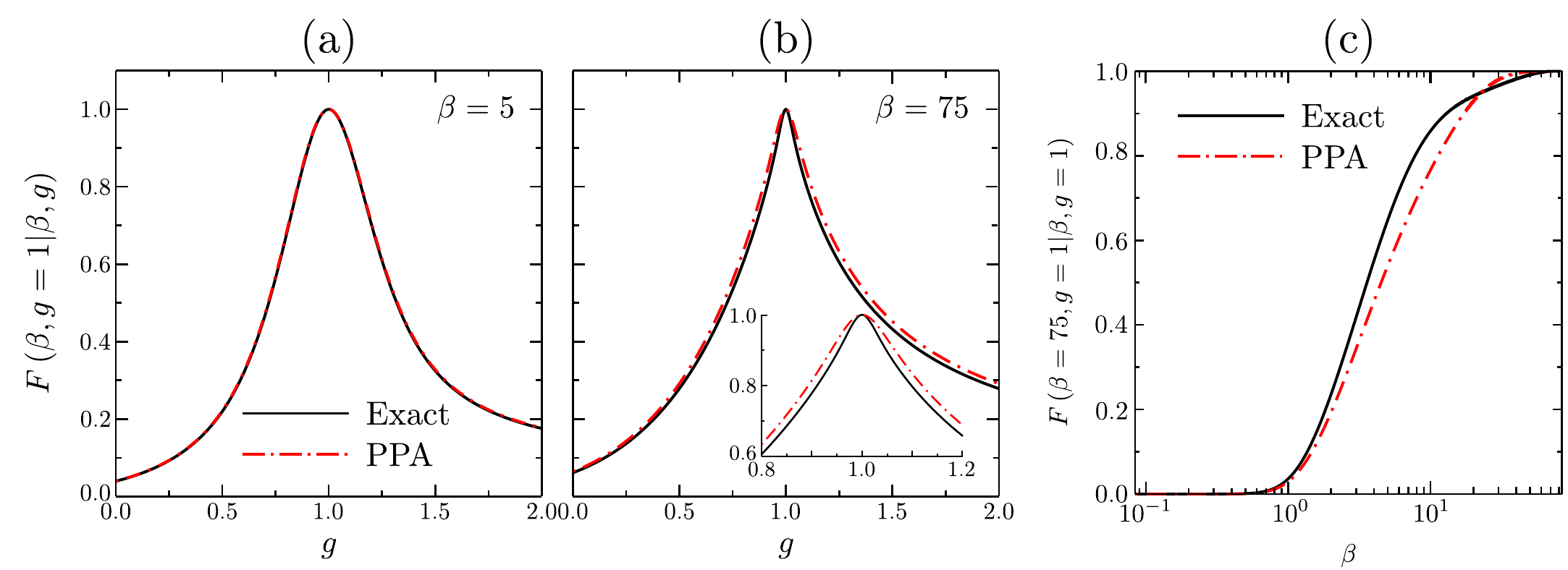}
	\caption{\label{UhlmannFide}Uhlmann fidelity between two Gibbs states differing by the value of the magnetic field or inverse temperature. In panels (a) and (b), the magnetic field of the first state is set to $g_c=1$ and that of the second state is the variable $g$.  Panel (a) shows the results in the high-temperature regime, in which the PPA approximation works very well. Panel (b) shows the results in the regime in which discrepancies between the exact expression and the PPA  are manifested (the inset zooms in close to the critical magnetic field). Panel (c) shows the Uhlmann fidelity between the Gibbs states $\rho\pap{\beta = 75,g=1}$ and  $\sigma\pap{\beta ,g=1}$ as a function of $\beta$. A logarithmic scale on the horizontal axis is used to capture a wide range of the inverse temperature. System size $N=50$. 
	} 
	\label{fig:Uhl_fid}
\end{figure}

We independently verified the correctness of the formula (\ref{eq:fidelity_full1}) by showing  that it reproduces the results obtained numerically by exact diagonalization for small system ($N=6,8,10$). For details of the numerical simulations, see~\ref{AppendixA}. 
\section{Numerical results for the fidelity susceptibility}\label{secChi}
In this section we  analyze the fidelity susceptibility with respect to external field $g$ for fixed inverse temperature $\beta$. We also set $\gamma=1$ (case of Ising model in transverse field). We thus focus on the quantity
\begin{equation}\label{eq:suscept_full}
\chi(\beta, g) = -\restr{\frac{d^2 }{d\delta^2}F\pap{\beta, \gamma=1, g\vert \beta, \gamma=1, g+\delta}}{\delta=0},
\end{equation}
which provides the leading nontrivial term in the expansion
\begin{equation}
F\pap{\beta,\gamma=1,g\vert\beta, \gamma=1, g+\delta} = 1 - \frac{1}{2} \chi(\beta, g) \delta^2 + \ldots
\end{equation}
By contrast, the  fidelity susceptibility obtained from the PPA reads
\begin{equation}
\label{eq:suscept_simple}
\chi_F^+(\beta,g) = -\restr{\frac{d^2 }{d\delta^2}F^+_F\pap{\beta, \gamma=1, g\vert \beta, \gamma=1, g+\delta}}{\delta=0} \quad \mathrm{(PPA)}. 
\end{equation}

\begin{figure}[t!]
\includegraphics[scale=1]{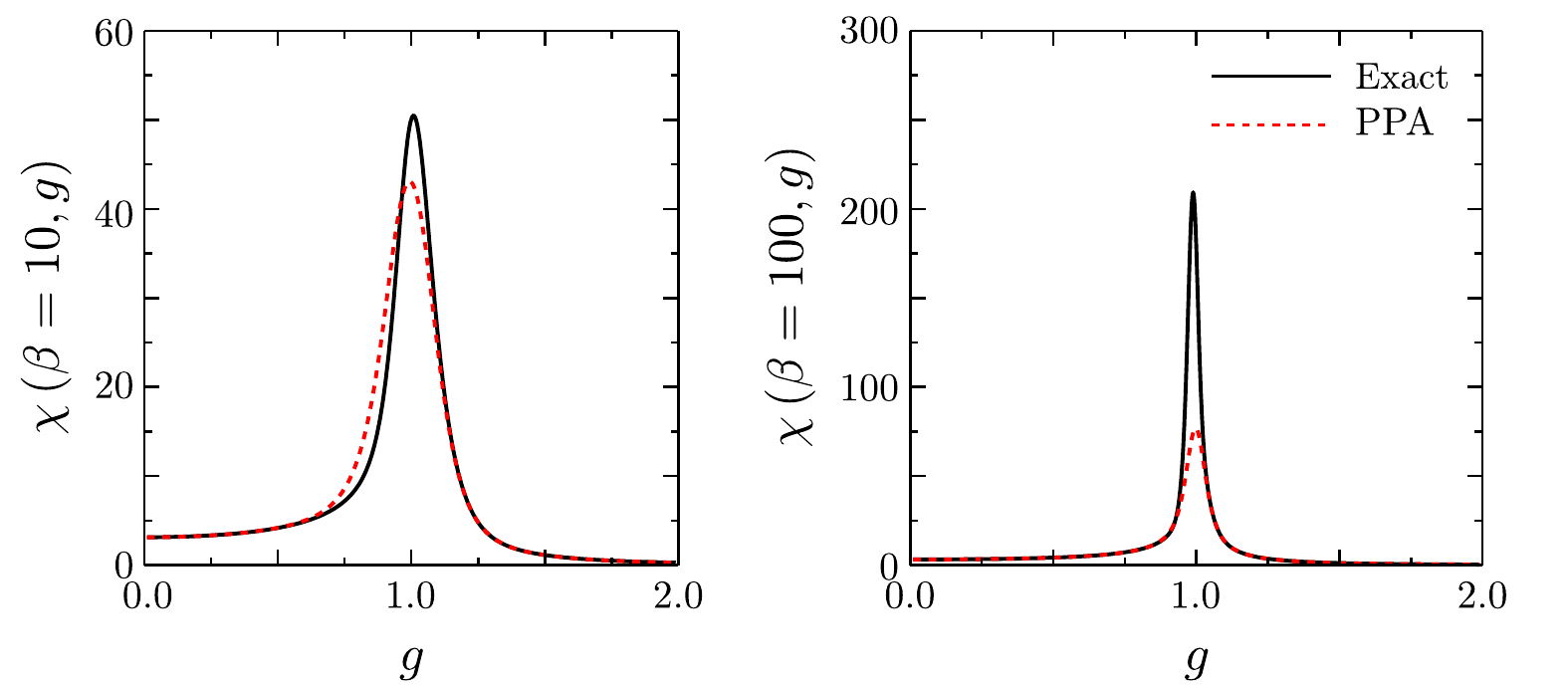}
\caption{\label{susc_g2}Fidelity susceptibility as a function of the transverse magnetic field $g$ at fixed temperature. The exact fidelity susceptibility, given by formulas~\eref{eq:fidelity_full1} and~\eref{eq:suscept_full} (black solid line), is compared with the corresponding PPA given by ~\eref{eq:fidelity_simp} and~\eref{eq:suscept_simple} (red dashed line) for a chain of $N=50$ spins. The location and magnitude of the maximum are altered in the PPA. The discrepancy between the exact and approximate results increases with $\beta$ as one approaches the low-temperature regime. } 
\end{figure}

A representative plot of the Uhlmann fidelity   is shown in Fig. \ref{fig:Uhl_fid} for  two Gibbs states
that differ in the value of magnetic field or 
inverse temperature. Significant discrepancies between the exact expression and the PPA are manifested at low temperatures. 
The choice of parameters has not been optimized to maximize the latter; discrepancies can be bigger, especially for general XY model ($\gamma \neq 1$), where states with negative parity can appear more often then for Ising model \cite{rams2015}. In the following, we examine the nature of the discrepancies by a detailed analysis of fidelity susceptibility.

\begin{figure} 
\includegraphics[scale=0.85]{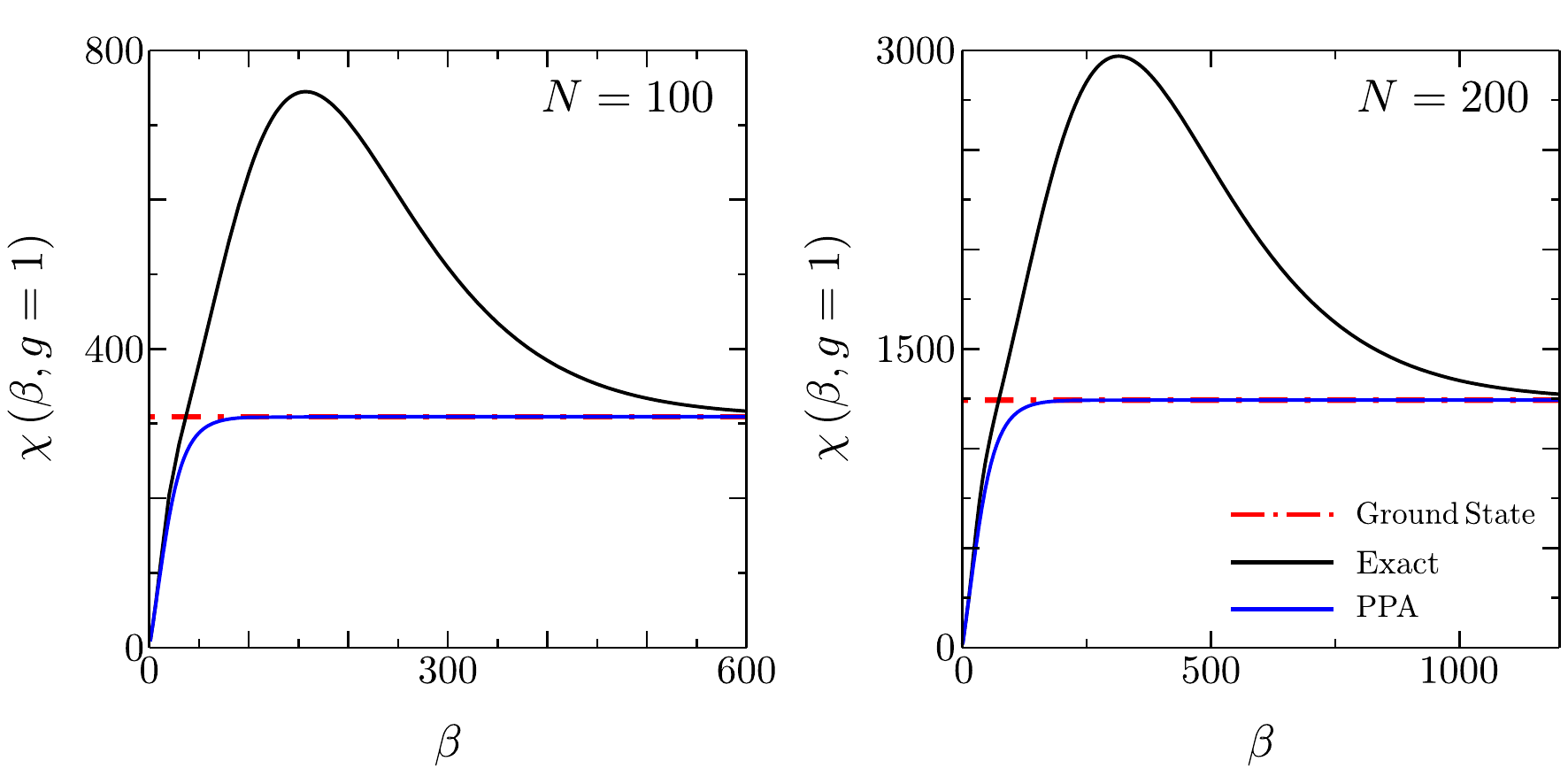}
\caption{Dependence of the maximum fidelity susceptibility  on inverse temperature at the critical point $g_c=1$. The exact expression (black solid line) computed using~\eref{eq:fidelity_full1} and~\eref{eq:suscept_full} is compared with the  PPA expression (red solid line) in ~\eref{eq:fidelity_simp} and~\eref{eq:suscept_simple}. The  blue dashed line corresponds to the ground-state fidelity susceptibility $\chi^+_{0}$~\eref{eq:ground_susc}. In the intermediate temperature regime, a significant enhancement of the fidelity susceptibility that scales as $N^2$ is found. The value of $\beta$ at the maximum scales linearly with system size.} 
\label{fig:suscept_max_beta}
\end{figure}

 In~\Fref{susc_g2}, we compare the exact and PPA results and show the dependence of fidelity susceptibility on the magnetic field for different $\beta$. Because the fidelity susceptibility is an even function of $g$, $\chi(\beta,g) = \chi(\beta, -g)$, we present the results only for $g>0$. The PPA yields qualitatively different results from the exact expression, with the difference being more pronounced in the low-temperature limit. Not only the value of the maximum differs, the PPA also leads to a shift of its location. It is clear that the discrepancy is enhanced in the vicinity of $g_c = 1$.   Figure \ref{fig:suscept_max_beta} shows the dependence of the fidelity susceptibility on $\beta$ at the critical point $g_c= 1$. It is found that  the enhancement of the exact fidelity susceptibility 
occurs for a specific value of $\beta$ that depends on the system size, a feature we analyze  next.

\subsection{Two-level Approximation of the Gibbs state}
Deviations between the PPA and the exact results are particularly pronounced at intermediate and low  temperatures. We next introduce an accurate  approximation  in this regime by truncating of the Gibbs state, taking only the two lowest energy states  into account. We refer to it as the the Two-Level Approximation (TLA) which yields the following truncation for the Gibbs state:
\begin{equation}\label{eq:approximation_def}
\rho_{\rm TLA}(\beta,g) = \frac{1}{Z_{\rm TLA}(\beta,g)}\pap{\mathrm{e}^{-\beta E_0^+(g)}\ket{g^{+}}\bra{g^{+}} 
+ \mathrm{e}^{-\beta E_0^-(g)}\ket{g^-}\bra{g^-}},
\end{equation}
where 
\begin{equation}\label{eq:Zapprox}
Z_{\rm TLA}(\beta,g) = \mathrm{e}^{-\beta E_0^+(g)} + \mathrm{e}^{-\beta E_0^-(g)}.
\end{equation}
Analogously, the neighboring thermal state can be written as 
\begin{eqnarray}
\rho_{\rm TLA}(\beta,g+\delta) = \frac{1}{Z_{\rm TLA}(\beta,g+\delta)}\Bigg( &\mathrm{e}^{-\beta E_0^+(g+\delta)}\ket{g+\delta^+}\bra{g+\delta^+}\nonumber\\ 
&+ \mathrm{e}^{-\beta E_0^-(g+\delta)}\ket{g+\delta^-}\bra{g+\delta^-}\Bigg).
\end{eqnarray}
In principle, the TLA should work well for big $\beta$, i.e., in the low-temperature regime. The accuracy of this approximation relies on the structure of energy levels and the relation between the dynamical and symmetry-breaking gap. Recall that the symmetry-breaking gap vanishes exponentially with $N$ for $g<1$ and is eight times smaller than the dynamical gap at the critical point. Because the energy $E_1^+$ of the first excited state in the positive-parity subspace is well separated from $E_0^+$ and $E_0^-$, one expects the contribution from $e^{-\beta E_1^+}$ and higher energy states to be small in comparison to~\Eref{eq:approximation_def}.

The Uhlmann fidelity between thermal states in the TLA  reads
\begin{eqnarray}
\label{eq:full_fidelity_approx}
F_{\rm TLA}\pap{\beta, g\vert \beta, g+\delta} = &\frac{1}{\sqrt{1+ \mathrm{e}^{-\beta \Delta(g)}}} \frac{F_0^+\pap{g\vert g+\delta}}{\sqrt{1+ \mathrm{e}^{-\beta \Delta(g+\delta)}}} +\nonumber\\ 
&+\frac{1}{\sqrt{1+ \mathrm{e}^{\beta \Delta(g)}}} \frac{F_0^-\pap{g\vert g+\delta}}{\sqrt{1+ \mathrm{e}^{\beta \Delta(g+\delta)}}} \quad \mathrm{(\rm TLA)},
\end{eqnarray}
where $\Delta(g)$ is a symmetry breaking gap defined in~(\ref{eq:symmetry_gap_definition}) and $F_0^+$ and $F_0^-$ denote the ground state fidelities in the even and odd parity subspace, respectively:
\begin{equation}
\label{eq:ground_susc_serie}
F^{\pm}_0\pap{g\vert g+\delta} = \vert \braket{g^{\pm} \vert g+\delta^{\pm}} \vert = 1 - \frac{\delta^2}{2} \chi^{\pm}_{0}(g) + \ldots
\end{equation}
The expression~\eref{eq:full_fidelity_approx} can be used as a starting point for computation of approximated fidelity susceptibility. Expressions for $\chi_0^{\pm}$ were derived analytically in~\cite{Damski2013}:
\begin{equation}
\chi_0^+(g ) = \frac{N^2}{16 g^2} \frac{g^N}{(g^N+1)^2} + \frac{N}{16g^2}\frac{g^N-g^2}{(g^N+1)(g^2-1)},
\end{equation}
\begin{equation}
\chi_0^-(g ) = -\frac{N^2}{16 g^2} \frac{g^N}{(g^N-1)^2} + \frac{N}{16g^2}\frac{g^N+g^2}{(g^N-1)(g^2-1)}.
\end{equation}
To explain the temperature dependence of the fidelity susceptibility,  we  focus on the case $g = 1$. Although for a finite system the maximum of the fidelity susceptibility is not exactly at $g_c=1$, it is very close to $1$ and in practice it is convenient to consider $\chi(\beta) = \chi(\beta, g=1)$. The ground-state positive and negative susceptibilities are given by
\begin{equation}
\label{eq:ground_susc}
\chi_0^+ = \frac{1}{32} N(N-1), \quad \chi_0^- = \frac{1}{96} (N^2-3N+2). 
\end{equation}
At $g=1$, we can use formulas~\eref{eq:full_fidelity_approx} and~\eref{eq:ground_susc_serie} to find $\chi(\beta)$ in the TLA
\begin{equation}
\label{eq:approximation_full}
\chi_{\rm TLA}(\beta) = \frac{\chi_0^+}{1+e^{-\frac{\beta\pi}{2N}}} + \frac{\chi_0^-}{1+e^{\frac{\beta\pi}{2N}}} + R(\beta), 
\end{equation}
where
\begin{figure}
\centering
\includegraphics[scale=1]{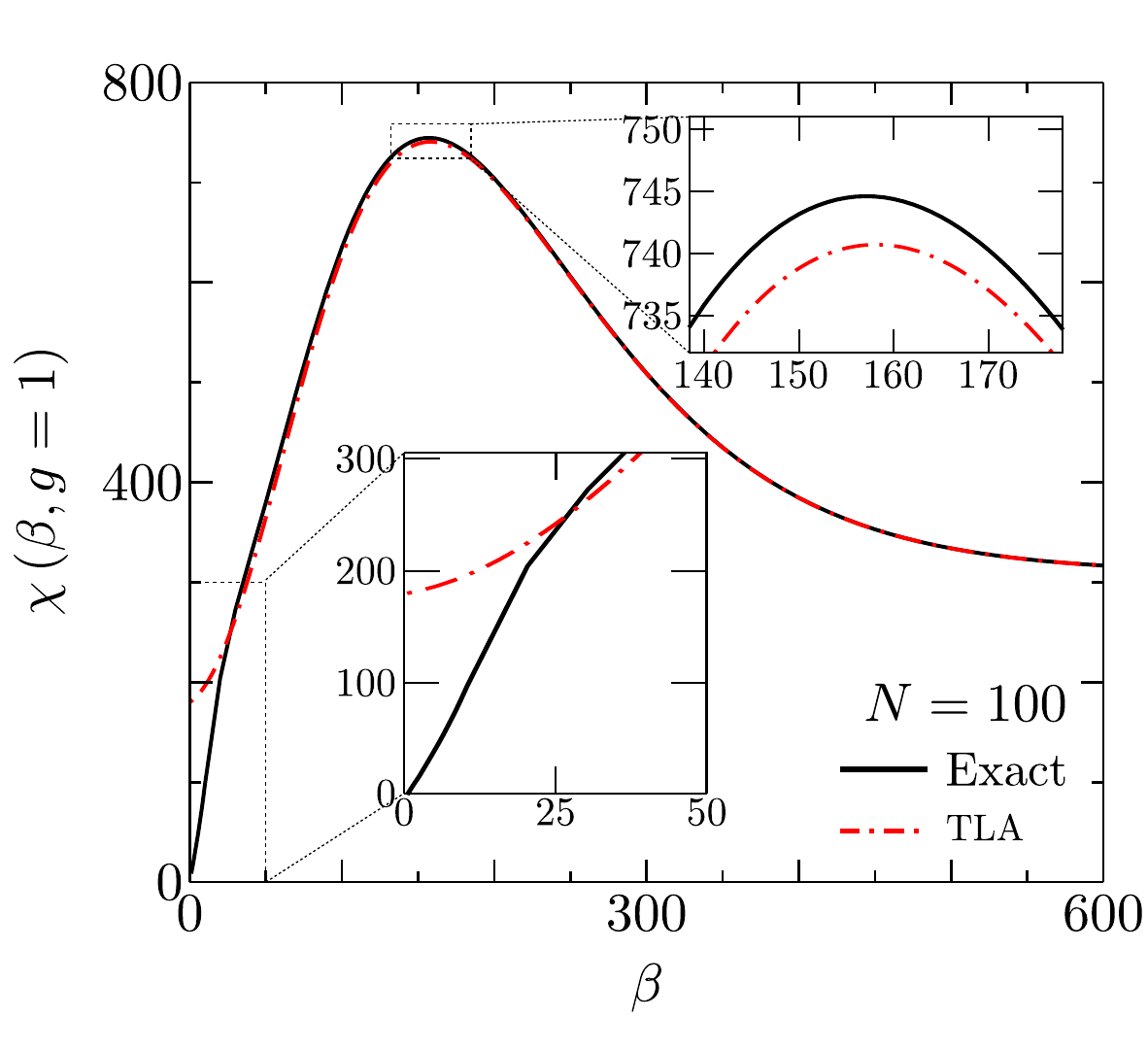}
\caption{\label{fig:approximation_check}Comparison between the maximal fidelity susceptibility computed with exact formula~\eref{eq:fidelity_full1} and the approximation~\eref{eq:approximation_full}. The match is excellent in a very wide range of temperatures and the relative error in the vicinity of maximum does not exceed 2\%.}
\end{figure}
\begin{equation}
R(\beta) = -\restr{\frac{d^2}{d\delta^2} \pap{\frac{1}{\sqrt{1+e^{-\frac{\beta \pi}{2N} }} \sqrt{1+e^{-\beta \Delta(1+\delta)}}} + \frac{1}{\sqrt{1+e^{\frac{\beta \pi}{2N} }} \sqrt{1+e^{\beta \Delta(1+\delta)}}} }}{\delta=0}.
\end{equation}
Here, we used the asymptotic behavior of the symmetry breaking gap at $g_c=1$; see Eqs.~\eref{eq:symmetry_gap_definition} and \eref{eq:symmetry_breaking_gap}.
Explicit evaluation yields
\begin{eqnarray}
R(\beta) = &
\frac{\beta ^2 e^{\frac{\beta\pi}{2N}} \Delta_c'^2}{2 \left(e^{\frac{\beta\pi}{2N}}+1\right)^{2}}
-\frac{3 \beta ^2 e^{\frac{\beta\pi}{2N}} \Delta_c'^2}{4 \left(e^{\frac{\beta\pi}{2N}}+1\right)^{3}}\nonumber\\
&+\frac{\beta ^2 e^{-\frac{\beta\pi}{2N}} \Delta_c'^2}{2 \left(e^{-\frac{\beta\pi}{2N}}+1\right)^{2}}
-\frac{3 \beta ^2 e^{-\frac{\beta\pi}{2N}} \Delta_c'^2}{4\left(e^{-\frac{\beta\pi}{2N}}+1\right)^{3}},
\end{eqnarray}
where the first derivative of the energy gap with respect to $g$ evaluated at $g_c=1$ is denoted by 
\begin{equation}
\Delta_c' = \restr{\frac{d}{d\delta}\Delta(1+\delta)}{\delta=0} = \restr{\frac{d}{dg}\Delta(g)}{g=1}.
\end{equation}
and  we note that terms depending on  the second derivative cancel out.
The explicit expression for $\Delta_c'$ readily follows using ~\eref{eq:energiesp1}  
\begin{equation}
\Delta_c' = \frac{1}{2} \Delta + 1 = \tan\pap{\frac{\pi}{4N}}+1 \approx \frac{\pi}{4N}+1.
\end{equation}
Substituting
\begin{equation}
x(\beta) = \frac{\pi}{4N}\beta
\end{equation}
and simplifying the expression for $R(\beta)$, one finds
\begin{equation}
R(x(\beta)) =\frac{x^2(\beta)}{16\cosh^2 \pas{x(\beta)}} \pap{\frac{16}{\pi^2}N^2 + \frac{8}{\pi} N +1}.
\end{equation}
From this expression it is clear that the maximum of the susceptibility scales quadratically with the system size. Moreover, the inverse temperature for which the maximum is achieved scales linearly with the system size. Further, the presence of the maximum  is due to the particular behavior of the first derivative of the symmetry breaking gap, 
\begin{equation}
\lim_{N\rightarrow \infty} \Delta_c' = 1 > 0.
\end{equation}
\Fref{fig:approximation_check} shows the range of $\beta$ where the approximation works. Obviously,  the approximation fails in the classical limit associated with large temperatures.   For $\beta \rightarrow 0$, both $\rho_{\rm Gibbs}(0,g)$ and $\rho_{\rm Gibbs}(0,g+\delta)$ are maximally mixed states, proportional to the identity operator. As a result, the fidelity 
$F(0,g \vert 0,g+\delta)=1$ for any $h$ and $\delta$. Consequently, the fidelity susceptibility vanishes by definition,
\begin{equation}
\label{lime}
\lim _{\beta\rightarrow 0}\, \chi_{\rm Exact}(\beta) = 0.
\end{equation} 
In this limit,  the TLA naturally fails 
and predicts incorrectly a finite value
\begin{equation}
\label{lima}
\lim _{\beta\rightarrow 0}\, \chi_{\rm TLA}(\beta) = \frac{1}{96} \pap{2N^2-3N+1},
\end{equation} 
which scales quadratically with the system size and diverges in the thermodynamic limit. 

The appearance of $\chi^{-}$ in \eref{eq:approximation_full} indicates the importance of the negative parity sector, which was omitted in previous studies \cite{Zanardi2007, Quan2009}.

\subsection{Thermal fidelity susceptibility and specific heat}
When the thermal states commute with each other, the Uhlmann fidelity is given in terms of the partition function, as we have discussed.
The Taylor series expansion of equation ~(\ref{UFcomm}) yields the thermal fidelity susceptibility, which is itself proportional to the specific heat of the system at constant magnetic field. Specifically, taking $\beta\rightarrow\beta-\delta\beta/2$ and $\beta'\rightarrow\beta+\delta\beta/2$, and expanding in $\delta\beta$, one find the leading term~\cite{Yu2007}
\begin{equation}
\xi(\beta,g)=-2\lim_{\delta\beta\rightarrow 0}\frac{\ln F(\beta-\delta\beta/2,g \vert \beta+\delta\beta/2,g)}{(\delta\beta)^2}=\frac{C_v(\beta,g)}{4\beta^2}.
\end{equation}
Next, we aim at computing the thermal fidelity susceptibility (and, thus, the specific heat) using the TLA~\eref{eq:approximation_def}. Exact expressions, in this case, can be calculated with help of equation~\eref{UFcomm} and knowledge of partition function~\eref{eq:FullZ}. Simplified expressions, used in literature, can be easily obtained just by substitution of the PPA partition function~\eref{eq:FermionicZ} into equation~\eref{UFcomm}. To derive the TLA thermal fidelity susceptibility we use the approximated formula for fidelity between equilibrium states with temperatures differing by $\delta$,
\begin{figure}[t!]
\includegraphics[width=1\linewidth]{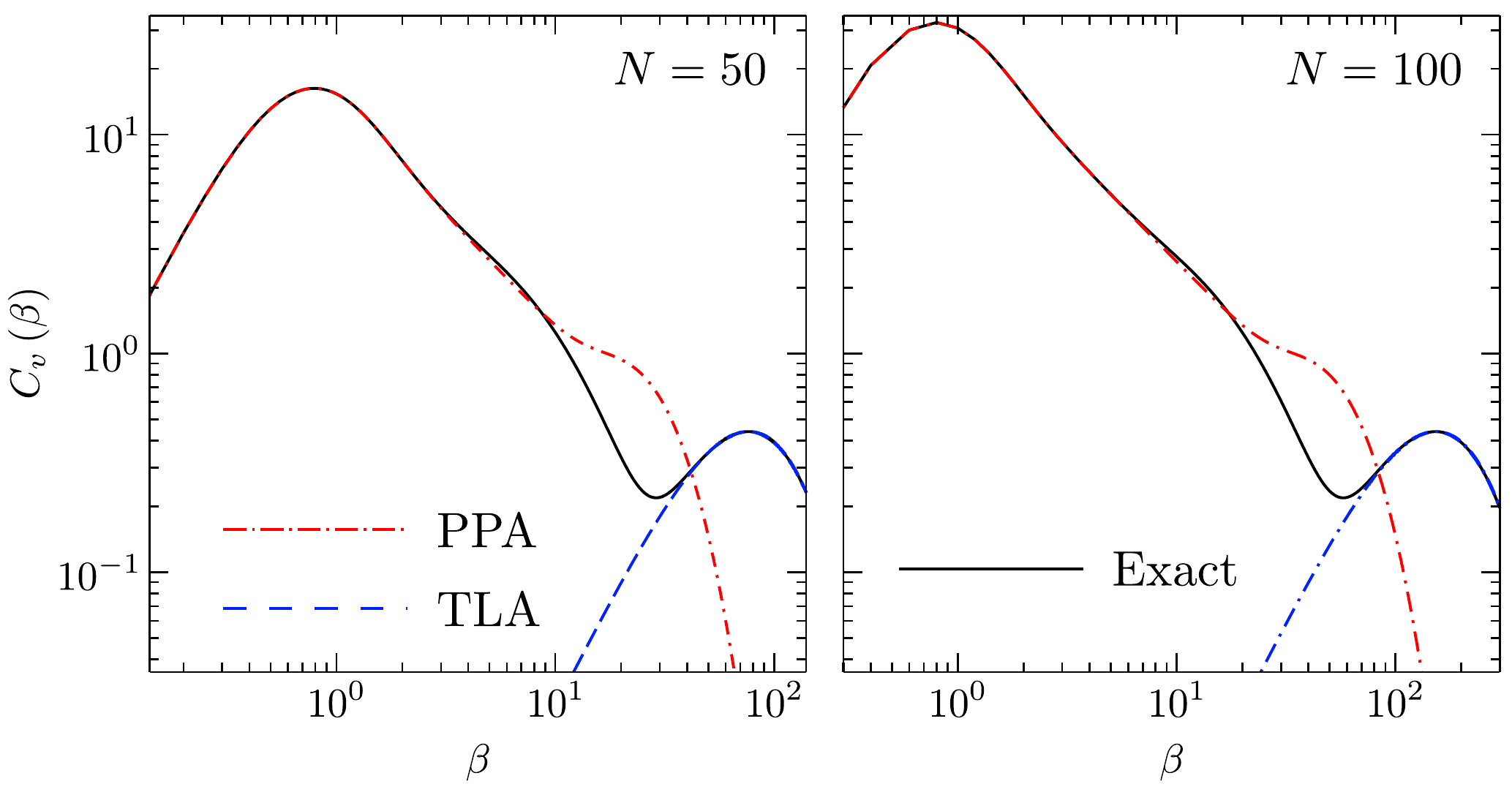}
\caption{\label{fig:thermal_suscept_50}Specific heat at $g=1$ computed with the exact formula for thermal susceptibility (black solid line) versus the corresponding PPA (red doted-dashed line). The TLA~\eref{eq:heat_approximation} is shown by the blue dashed line. The chosen system sizes are $N=50$ and $N=100$. Note that in the low-temperature regime the height of the maximum is the same for $N=100$ and $N=50$. To bring out the characteristic features of both temperature regimes, a  logarithmic scale is used in both axes.}
\end{figure}
\begin{eqnarray}
F_{\rm TLA}\pap{\beta,g\vert\beta+\delta,g} &= F\pap{\beta, \gamma=1, g\vert \beta+\delta,\gamma=1 ,g}\nonumber \\
&= \frac{Z_{\rm TLA}(\beta+\delta/2)}{\sqrt{Z_{\rm TLA}(\beta,g)\, Z_{\rm TLA}(\beta+\delta,g)}},
\end{eqnarray} 
where $Z_{\rm TLA}$ is given by formula~\eref{eq:Zapprox}. 
The explicit expression reads
\begin{equation}
F_{\rm TLA}\pap{\beta,g\vert\beta+\delta, g} = \frac{1+ e^{(-\beta+\delta/2) \Delta(g)}}{\sqrt{1+e^{-\beta \Delta(g)}}\sqrt{1+e^{(\beta+\delta) \Delta(g)}}}, 
\end{equation}
where $\Delta(g)$ is symmetry-breaking gap defined in the formula~\eref{eq:symmetry_gap_definition}. Using it, the calculation of specific heat yields a remarkably simple expression
\begin{eqnarray}
C_{v}\pap{\beta,g} = 4\beta^2\, \xi(\beta,g) &= -4 \beta^2\, \frac{\partial^2 }{\partial \delta^2}F(\beta,g\vert\beta+\delta,g) \vert_{\delta=0}\nonumber\\
&= \frac{\beta^2 \Delta^2(g)}{4}{\rm sech}^2\pas{\frac{\beta \Delta(g)}{2}}.\quad {\rm (TLA)}
\end{eqnarray}
Above formula gives TLA approximation for arbitrary values of $\beta$ and $g$, one expects that approximation works better for larger $\beta$. In order to channel the discussion, we restrict ourselves to the case $g=1$. Then, substituting $x(\beta) = \frac{\beta \pi}{4N}$, one finds
\begin{equation}\label{eq:heat_approximation}
C_{v} \pap{\beta}= C_v(\beta,g=1) = x^2(\beta){\rm sech}^{2}\pas{ x(\beta)}.\quad {\rm (TLA)}
\end{equation}
The temperature dependence of the specific heat at the critical point is shown in the \Fref{fig:thermal_suscept_50}, where the exact results are compared with the PPA and TLA. A characteristic feature of the temperature dependence of the specific heat is the big contrast between the high- and low-temperature regimes. In the high-temperature regime, the specific heat exhibits a sharp peak, which scales linearly with the system size. On the other hand, at low temperatures, the TLA  formula~\eref{eq:heat_approximation} provides an accurate prediction, with a maximum $C_{v}^{\max} \approx 0.439229$ independent of the system size. By contrast, the PPA  partition function gives results very close to the exact one for high temperatures, as in the case of the susceptibility related to the field $g$ (compare with the previous section). 
In short, the exact results are well reproduced by the TLA in the low-temperature regime, and by the PPA in the high-temperature regime.

\section{Conclusions}
An important family of paradigmatic spin models-including the one-dimensional XY and Ising models-are integrable and can be expressed in terms of free fermions. The parity operator commutes with the Hamiltonian and it is often convenient to simplify their description by considering only the positive-parity subspace. This yields simple approximate formulae for relevant quantities such as the partition function which are ubiquitous in the literature~\cite{sachdev_2011,Suzuki_2013}.

We have shown that such an approach fails in the characterization of quantum critical phenomena using the ground-state fidelity and fidelity susceptibility. Using an algebraic approach to exactly account for the complete Hilbert space~\cite{Bialonczyk2020}, we have shown that the exact result for the fidelity susceptibility between thermal states qualitatively differs from the conventional approximation, away from the classical high-temperature regime. The discrepancy is pronounced in the quantum, low-temperature limit when the accurate treatment of the low-lying energy states is crucial. Furthermore, the discrepancy is robust against variations of the system size and is manifested even in the thermodynamic limit.

Our results show the limitation of disregarding the odd parity subspace in the characterization of integrable spin chains at finite temperature and are potentially relevant to applications ranging from quantum thermodynamics to parameter estimation, among other quantum technologies using quantum critical spin chains. The exact expressions we have provided for the fidelity between thermal states should be of broad interest in the quantum-information characterization of these systems at finite temperature, with applications ranging from quantum thermodynamics to many-body and criticality-enhanced quantum metrology \cite{Rams18,Frerot18,Chabuda2020}. Likewise, the exact treatment of the fidelity susceptibility is required for the analysis of their critical phenomena and could be used for benchmarking quantum simulators, annealing devices, and the performance of quantum algorithms using quantum spin chains as a testbed.

Our results apply to the canonical Gibbs state at thermal equilibrium resulting from thermalization dynamics without other conserved quantities than the energy in chains of fixed size. An interesting outlook concerns the characterization of the fidelity and fidelity susceptibility in the Generalized Gibbs Ensemble, i.e., in the presence of additional invariants of motion.

\section*{Acknowledgments}
M. B. would like to thank Bogdan Damski and Andrzej Syrwid for insightful discussions. We further thank Marek M. Rams for feedback on the manuscript. M. B. also acknowledges the support of the Polish National Science Center under scholarship ETIUDA (2020/36/T/ST3/00332) and research grant DEC-2016/23/B/ST3/01152. F.J.~G-R thanks the University of Luxembourg for hospitality during the completion of this work. This work is further supported by the Spanish Ministerio de Ciencia e Innovaci\'on (PID2019-109007GA-I00).
\appendix
\section{Details of numerical simulations}\label{AppendixA}
Numerical simulations were performed with \emph{Wolfram Mathematica}, version 12. Fidelity susceptibility from the formula~\eref{eq:FullZ} is calculated first by symbolic differentiation to obtain analytical formula and then insertion of the parameters $\beta$ and $g$ - facilitating a very high precision. In benchmarking the analytical expressions with the the numerically -exact diagonalization, it is required  to approximate  the second derivative. We use central finite difference. In order to achieve satisfactory precision we have to use the fourth order  according to which  the second derivative of arbitrary function $f(x)$ is given by~\cite{Hassan2012}
\begin{eqnarray}
\frac{d^2}{dx^2} f(x) \approx \frac{1}{\Delta x^2}\Bigg(&-\frac{1}{12} f(x-2\Delta x) + \frac{4}{3} f(x-\Delta x)- \frac{5}{2} f(x)\nonumber\\
& + \frac{4}{3} f(x+\Delta x) - \frac{1}{12} f(x+2\Delta x)\Bigg).
\end{eqnarray}
This formula can be obtained by interpolating the function $f$ with a polynomial of degree four and can be proven to approximate the second derivative in $\Delta x^4$ order \cite{Hassan2012}. The fidelity in exact diagonalization is computed directly from the definition with help of built-in \emph{Mathematica} functions \texttt{MatrixExp} and \texttt{MatrixPower}. 

\begin{figure}[t!]
\includegraphics[scale=1]{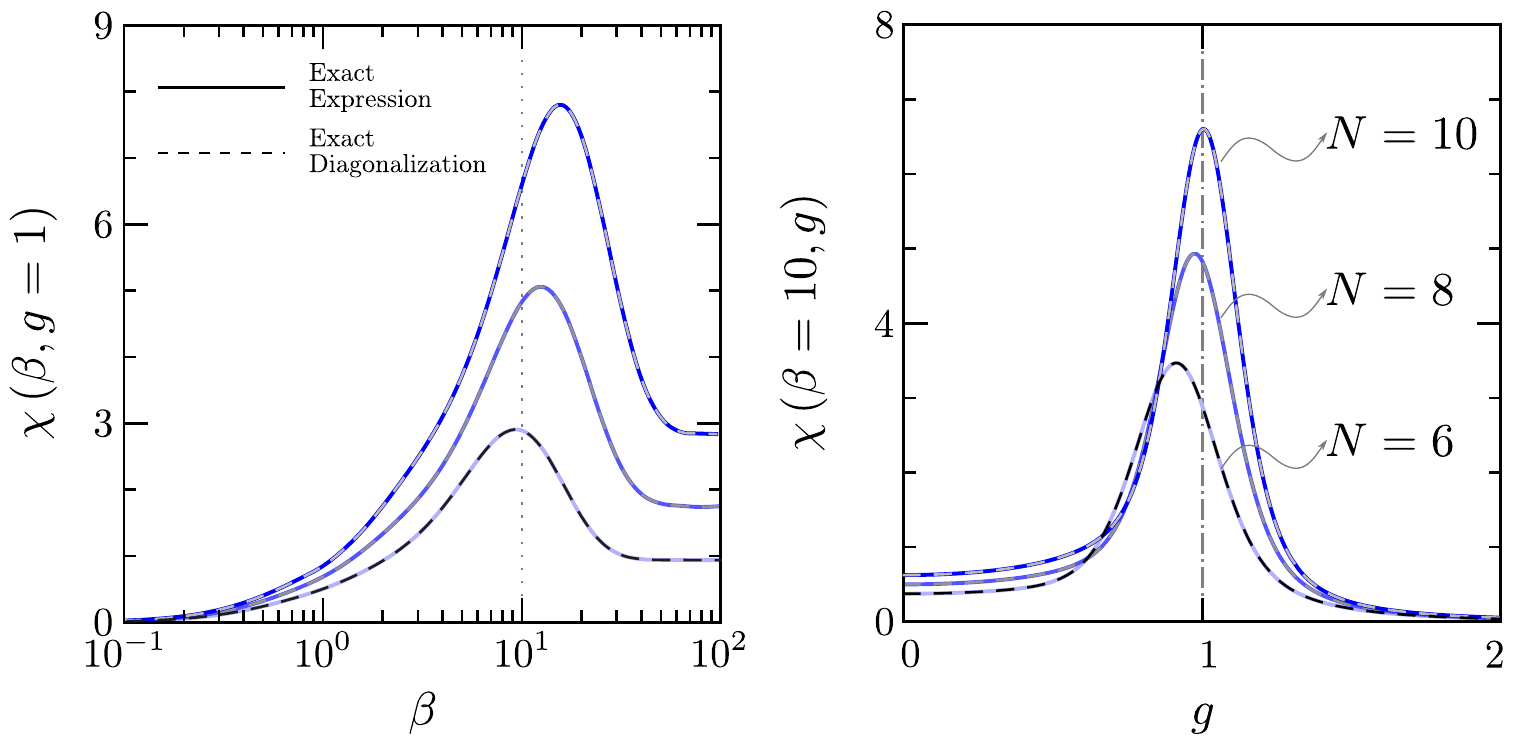}
\caption{\label{fig:check_with_exact} The exact analytical expression for the fidelity susceptibility given by equation (\ref{eq:fidelity_full1}), using the formula for the total partition function (Eq.~(\ref{eq:FullZ})) reproduces the results obtained by the numerically-exact diagonalization. In the left panel, the dotted line corresponds at $\beta=10$, this value is used in the right panel to plot the fidelity susceptibility as a function of the magnetic field.
}
\end{figure}

Figure~\ref{fig:check_with_exact} shows that  the numerically-exact results match the exact analytical expressions. Moreover, plots for different system sizes  $N$ indicate that the position of the maximum approaches the thermodynamic value at $g_c=1$ already for $N=10$ (see also figure \ref{susc_g2}). The height of the maximum grows with the system size like $N^2$, as shown in section \ref{secChi}.

\newpage 
\section*{References}
\bibliographystyle{iopart-num}
\bibliography{references}

\providecommand{\newblock}{}
\begin{thebibliography}{10}
\expandafter\ifx\csname url\endcsname\relax
  \def\url#1{{\tt #1}}\fi
\expandafter\ifx\csname urlprefix\endcsname\relax\def\urlprefix{URL }\fi
\providecommand{\eprint}[2][]{\url{#2}}

\bibitem{Amico2008}
Amico L, Fazio R, Osterloh A and Vedral V 2008 {\em Rev. Mod. Phys.\/} {\bf
  80}(2) 517--576
  \urlprefix\url{https://link.aps.org/doi/10.1103/RevModPhys.80.517}

\bibitem{sachdev_2011}
Sachdev S 2011 {\em Quantum Phase Transitions\/} 2nd ed (Cambridge University
  Press)

\bibitem{Suzuki_2013}
Suzuki S, ichi Inoue J and Chakrabarti B~K 2013 {\em Quantum Ising Phases and
  Transitions in Transverse Ising Models\/} 2nd ed (Springer)

\bibitem{information_transitions}
Casta{\~{n}}os O, L{\'{o}}pez-Pe{\~{n}}a R, Nahmad-Achar E and Hirsch J~G 2012
  {\em Journal of Physics: Conference Series\/} {\bf 403} 012003
  \urlprefix\url{https://doi.org/10.1088/1742-6596/403/1/012003}

\bibitem{Quan2006}
Quan H~T, Song Z, Liu X~F, Zanardi P and Sun C~P 2006 {\em Phys. Rev. Lett.\/}
  {\bf 96}(14) 140604
  \urlprefix\url{https://link.aps.org/doi/10.1103/PhysRevLett.96.140604}

\bibitem{ViyuelaPRB18}
Mera B, Vlachou C, Paunkovi\ifmmode~\acute{c}\else \'{c}\fi{} N, Vieira V~R and
  Viyuela O 2018 {\em Phys. Rev. B\/} {\bf 97}(9) 094110
  \urlprefix\url{https://link.aps.org/doi/10.1103/PhysRevB.97.094110}

\bibitem{Zanardi2006}
Zanardi P and Paunkovi\ifmmode~\acute{c}\else \'{c}\fi{} N 2006 {\em Phys. Rev.
  E\/} {\bf 74}(3) 031123
  \urlprefix\url{https://link.aps.org/doi/10.1103/PhysRevE.74.031123}

\bibitem{Gu2008}
Gu S~J 2010 {\em Inter. J. Modern Phys. B\/} {\bf 24} 4371--4458
  \urlprefix\url{https://doi.org/10.1142/S0217979210056335}

\bibitem{Yu2007}
You W~L, Li Y~W and Gu S~J 2007 {\em Phys. Rev. E\/} {\bf 76}(2) 022101
  \urlprefix\url{https://link.aps.org/doi/10.1103/PhysRevE.76.022101}

\bibitem{zanardi_geometric}
Zanardi P, Giorda P and Cozzini M 2007 {\em Phys. Rev. Lett.\/} {\bf 99}(10)
  100603 \urlprefix\url{https://link.aps.org/doi/10.1103/PhysRevLett.99.100603}

\bibitem{Scherer_2009}
Scherer D~D, Müller C~A and Kastner M 2009 {\em Journal of Physics A:
  Mathematical and Theoretical\/} {\bf 42} 465304
  \urlprefix\url{https://doi.org/10.1088/1751-8113/42/46/465304}

\bibitem{Lacki_2013}
\L{}acki M, Damski B and Zakrzewski J 2014 {\em Phys. Rev. A\/} {\bf 89}(3)
  033625 \urlprefix\url{https://link.aps.org/doi/10.1103/PhysRevA.89.033625}

\bibitem{Albuquerque10}
Albuquerque A~F, Alet F, Sire C and Capponi S 2010 {\em Phys. Rev. B\/} {\bf
  81}(6) 064418
  \urlprefix\url{https://link.aps.org/doi/10.1103/PhysRevB.81.064418}

\bibitem{Wang15}
Wang L, Liu Y~H, Imri\ifmmode~\check{s}\else \v{s}\fi{}ka J, Ma P~N and Troyer
  M 2015 {\em Phys. Rev. X\/} {\bf 5}(3) 031007
  \urlprefix\url{https://link.aps.org/doi/10.1103/PhysRevX.5.031007}

\bibitem{Chabuda2020}
Chabuda K, Dziarmaga J, Osborne T~J and Demkowicz-Dobrza{\'{n}}ski R 2020 {\em
  Nature Communications\/} {\bf 11} 250 ISSN 2041-1723
  \urlprefix\url{https://doi.org/10.1038/s41467-019-13735-9}

\bibitem{Islam2011}
Islam R, Edwards E~E, Kim K, Korenblit S, Noh C, Carmichael H, Lin G~D, Duan
  L~M, Joseph~Wang C~C, Freericks J~K and Monroe C 2011 {\em Nature
  Communications\/} {\bf 2} 377 ISSN 2041-1723
  \urlprefix\url{https://doi.org/10.1038/ncomms1374}

\bibitem{Zhang2017}
Zhang J, Pagano G, Hess P~W, Kyprianidis A, Becker P, Kaplan H, Gorshkov A~V,
  Gong Z~X and Monroe C 2017 {\em Nature\/} {\bf 551} 601--604 ISSN 1476-4687
  \urlprefix\url{https://doi.org/10.1038/nature24654}

\bibitem{Zanardi2007}
Zanardi P, Quan H~T, Wang X and Sun C~P 2007 {\em Phys. Rev. A\/} {\bf 75}(3)
  032109 \urlprefix\url{https://link.aps.org/doi/10.1103/PhysRevA.75.032109}

\bibitem{Quan2009}
Quan H~T and Cucchietti F~M 2009 {\em Phys. Rev. E\/} {\bf 79}(3) 031101
  \urlprefix\url{https://link.aps.org/doi/10.1103/PhysRevE.79.031101}

\bibitem{jozsa_fidelity}
Jozsa R 1994 {\em Journal of Modern Optics\/} {\bf 41} 2315--2323
  \urlprefix\url{https://doi.org/10.1080/09500349414552171}

\bibitem{QingPRE18}
Luo Q, Zhao J and Wang X 2018 {\em Phys. Rev. E\/} {\bf 98}(2) 022106
  \urlprefix\url{https://link.aps.org/doi/10.1103/PhysRevE.98.022106}

\bibitem{BrunoPRL17}
Mera B, Vlachou C, Paunkovi\ifmmode~\acute{c}\else \'{c}\fi{} N and Vieira V~R
  2017 {\em Phys. Rev. Lett.\/} {\bf 119}(1) 015702
  \urlprefix\url{https://link.aps.org/doi/10.1103/PhysRevLett.119.015702}

\bibitem{Katsura62}
Katsura S 1962 {\em Phys. Rev.\/} {\bf 127}(5) 1508--1518
  \urlprefix\url{https://link.aps.org/doi/10.1103/PhysRev.127.1508}

\bibitem{Kapitonov1998}
Kapitonov V~S and Il'inskii K~N 1998 {\em Journal of Mathematical Sciences\/}
  {\bf 88} 233--243 ISSN 1573-8795
  \urlprefix\url{https://doi.org/10.1007/BF02364984}

\bibitem{Bialonczyk2020}
Bia{\l}o{\'{n}}czyk M, G\'omez-Ruiz F~J and del Campo A 2021 Exact thermal
  properties of integrable spin chains (\textit{Preprint} \eprint{2103.16322})

\bibitem{Damski2014}
Damski B and Rams M~M 2013 {\em Journal of Physics A: Mathematical and
  Theoretical\/} {\bf 47} 025303
  \urlprefix\url{https://doi.org/10.1088/1751-8113/47/2/025303}

\bibitem{Damski_2011}
Rams M~M and Damski B 2011 {\em Phys. Rev. A\/} {\bf 84}(3) 032324
  \urlprefix\url{https://link.aps.org/doi/10.1103/PhysRevA.84.032324}

\bibitem{Damski2013}
Damski B 2013 {\em Phys. Rev. E\/} {\bf 87}(5) 052131
  \urlprefix\url{https://link.aps.org/doi/10.1103/PhysRevE.87.052131}

\bibitem{ExtendedXY}
Zhang G and Song Z 2015 {\em Phys. Rev. Lett.\/} {\bf 115}(17) 177204
  \urlprefix\url{https://link.aps.org/doi/10.1103/PhysRevLett.115.177204}

\bibitem{Lieb1961}
Lieb E, Schultz T and Mattis D 1961 {\em Annals of Physics\/} {\bf 16} 407 --
  466 ISSN 0003-4916
  \urlprefix\url{http://www.sciencedirect.com/science/article/pii/0003491661901154}

\bibitem{rams2015}
Okuyama M, Yamanaka Y, Nishimori H and Rams M~M 2015 {\em Phys. Rev. E\/} {\bf
  92}(5) 052116
  \urlprefix\url{https://link.aps.org/doi/10.1103/PhysRevE.92.052116}

\bibitem{BialonczykSymmetry}
Bia{\l}o{\'{n}}czyk M and Damski B 2020 {\em Journal of Statistical Mechanics:
  Theory and Experiment\/} {\bf 2020} 013108
  \urlprefix\url{https://doi.org/10.1088/1742-5468/ab609a}

\bibitem{Carbera1987}
Cabrera G~G and Jullien R 1987 {\em Phys. Rev. B\/} {\bf 35}(13) 7062--7072
  \urlprefix\url{https://link.aps.org/doi/10.1103/PhysRevB.35.7062}

\bibitem{ningWu2020}
Wu N 2020 {\em Phys. Rev. E\/} {\bf 101}(4) 042108
  \urlprefix\url{https://link.aps.org/doi/10.1103/PhysRevE.101.042108}

\bibitem{Pfeuty70}
Pfeuty P 1970 {\em Annals of Physics\/} {\bf 57} 79--90 ISSN 0003-4916
  \urlprefix\url{https://www.sciencedirect.com/science/article/pii/0003491670902708}

\bibitem{Barouch1975}
Barouch E, McCoy B~M and Dresden M 1970 {\em Phys. Rev. A\/} {\bf 2}(3)
  1075--1092 \urlprefix\url{https://link.aps.org/doi/10.1103/PhysRevA.2.1075}

\bibitem{Deng2011}
Deng S, Ortiz G and Viola L 2011 {\em Phys. Rev. B\/} {\bf 83}(9) 094304
  \urlprefix\url{https://link.aps.org/doi/10.1103/PhysRevB.83.094304}

\bibitem{Fei19}
Fei Z and Quan H~T 2019 {\em Phys. Rev. Research\/} {\bf 1}(3) 033175
  \urlprefix\url{https://link.aps.org/doi/10.1103/PhysRevResearch.1.033175}

\bibitem{Uhlmann76}
Uhlmann A 1976 {\em Reports on Mathematical Physics\/} {\bf 9} 273 -- 279 ISSN
  0034-4877
  \urlprefix\url{http://www.sciencedirect.com/science/article/pii/0034487776900604}

\bibitem{Rams18}
Rams M~M, Sierant P, Dutta O, Horodecki P and Zakrzewski J 2018 {\em Phys. Rev.
  X\/} {\bf 8}(2) 021022
  \urlprefix\url{https://link.aps.org/doi/10.1103/PhysRevX.8.021022}

\bibitem{Frerot18}
Fr\'erot I and Roscilde T 2018 {\em Phys. Rev. Lett.\/} {\bf 121}(2) 020402
  \urlprefix\url{https://link.aps.org/doi/10.1103/PhysRevLett.121.020402}

\bibitem{Hassan2012}
Hassan H, Mohamad A and Atteia G 2012 {\em Journal of Computational and Applied
  Mathematics\/} {\bf 236} 2622 -- 2631 ISSN 0377-0427
  \urlprefix\url{http://www.sciencedirect.com/science/article/pii/S0377042711006339}

\end{thebibliography}

\end{document}